\documentclass[11pt]{article}

\usepackage{lipsum} % Package to generate dummy text throughout this template

\usepackage{bm}
\pdfoutput=1
\usepackage{amsmath,amsfonts,amsthm}
\usepackage{esdiff}  
\usepackage{booktabs}  % For better-looking tables
\usepackage{url}  % To make clickable url's
\usepackage[skip=1.5pt,font=small]{caption}

\usepackage{caption}
\usepackage{bbm}

\usepackage[usenames,dvipsnames,svgnames,table]{xcolor}

\usepackage{graphicx}
\usepackage{hyperref}  % for footnotes?
\usepackage{cleveref}  % Detects type of ref when you write \cref. 
	\crefname{equation}{equation}{equations}
	\crefname{figure}{figure}{figures}	
	\crefname{table}{table}{tables}

\usepackage{tikz}  % To draw h(circle)

\usepackage{cite}  % In order to have the citations go: [1-4] instead of [1][2][3][4].

\usepackage[sc]{mathpazo} % Use the Palatino font
\usepackage[T1]{fontenc} % Use 8-bit encoding that has 256 glyphs
\usepackage{microtype} % Slightly tweak font spacing for aesthetics

\usepackage{multicol} % Used for the two-column layout of the document
\usepackage[margin={1cm,1.5cm}]{geometry}
\usepackage{booktabs} % Horizontal rules in tables
\usepackage{float} % Required for tables and figures in the multi-column environment - they need to be placed in specific locations with the [H] (e.g. \begin{table}[H])
\usepackage{hyperref} % For hyperlinks in the PDF

\usepackage{lettrine} % The lettrine is the first enlarged letter at the beginning of the text
\usepackage{paralist} % Used for the compactitem environment which makes bullet points with less space between them

\usepackage{abstract} % Allows abstract customization
\usepackage{titlesec} % Allows customization of titles
\renewcommand\thesection{\Roman{section}} % Roman numerals for the sections
\renewcommand\thesubsection{\Alph{subsection}} % Alphabet for subsections
\titleformat{\section}[block]{\large\scshape\centering\bfseries}{\thesection.}{1em}{} % Change the look of the section titles

\titleformat{\subsection}[block]{\scshape\centering}{\thesubsection.}{1em}{} % Change the look of the section titles

\usepackage{fancyhdr} % Headers and footers
\DeclareCaptionFormat{myformat}{#1#2#3\hrulefill}
\usepackage[tableposition=top]{caption}

\usepackage{authblk}
\title{\vspace{-15mm}\fontsize{16pt}{16pt}\selectfont\textbf{Statistical data assimilation for estimating electrophysiology simultaneously with connectivity within a biological neuronal network}} % 
\author[1,2]{Eve Armstrong\thanks{earmst01@nyit.edu}}
\affil[1]{Department of Physics, New York Institute of Technology, New York, NY 10023, USA}
\affil[2]{Department of Astrophysics, American Museum of Natural History, New York, NY 10024, USA}
\date{(Dated: \today)}
\begin{document}
\maketitle % Insert title
%\thispagestyle{fancy} % All pages have headers and footers
%----------------------------------------------------------------------------------------
%	ABSTRACT
%----------------------------------------------------------------------------------------

\begin{abstract}
\noindent
A method of data assimilation (DA) is employed to estimate electrophysiological parameters of neurons simultaneously with their synaptic connectivity in a small model biological network.  The DA procedure is cast as an optimization, with a cost function consisting of both a measurement error and a model error term.  An iterative reweighting of these terms permits a systematic method to identify the lowest minimum, within a local region of state space, on the surface of a non-convex cost function.  In the model, two sets of parameter values are associated with two particular functional modes of network activity: simultaneous firing of all neurons, and a pattern-generating mode wherein the neurons burst in sequence.  The DA procedure is able to recover these modes if: i) the stimulating electrical currents have chaotic waveforms, and ii) the measurements consist of the membrane voltages of all neurons in the circuit.  Further, this method is able to prune a model of unnecessarily high dimensionality to a representation that contains the maximum dimensionality required to reproduce the provided measurements.  This paper offers a proof-of-concept that DA has the potential to inform laboratory designs for estimating properties in small and isolatable functional circuits.
\end{abstract}

\begin{multicols}{2}
\section{INTRODUCTION}

Biological circuits can generate patterned electrical outputs that manifest in rhythmic motor behaviors vital for survival, such as respiration and heartbeat.  The means by which neurons within such a circuit act in coordination to yield reliable output is a largely open problem.  Progress in this area has been made chiefly via the examination of small (fewer than ten-cell) networks that are functional even when isolated from the animal~\cite{marder1996principles,kristan2005neuronal,marder2005invertebrate,mulloney2007local,marder2007tightly,smarandache2009coordination,grashow2009reliable,grashow2010compensation,turrigiano2011too,marder2016complicating,gunaratne2017variations}.  Through this work, it has been shown that the relationships among cellular and synapse parameters are fiercely nonlinear.  For example, couplings have been identified between two particular parameters of a circuit, where a change in the value of one parameter is associated with a change in the value of the second, so that the circuit output is maintained (e.g. Ref~\cite{grashow2010compensation}).  One major obstacle to dissecting these relationships is the difficulty of measuring more than just a few parameters \textit{simultaneously}.

This paper examines the potential of statistical data assimilation (DA) to simultaneously estimate multiple parameter values in a small biological network model, via measurements that are currently obtainable in a laboratory.  DA - or \lq\lq optimal estimation\rq\rq\ in control theory - is an inverse formulation~\cite{tarantola2005inverse}: calculating from measurements the processes that produced those measurements, given a model's response to input.  DA is distinct from machine learning in that a dynamical physical model is assumed to give rise to any measured quantities.  The model may contain state variables that are unmeasured and parameters that are unknown.  Invented for numerical weather prediction~\cite{kimura2002numerical,kalnay2003atmospheric,evensen2009data,betts2010practical,whartenby2013number,an2017estimating}, DA is designed to ascertain what measurements of a system are required for accurate state and parameter estimation of the associated model.  That is, DA can identify which measurements contain sufficient information about \textit{unmeasured} quantities to permit the completion of a model.  The test of an estimate's success is the predictive power of that completed model.  

Recently, DA has been applied to single-neuron models in  neuroscience~\cite{schiff2009kalman,toth2011dynamical,kostuk2012dynamical,hamilton2013real,meliza2014estimating,nogaret2016automatic}.  The specific formulation employed in this paper has been tested with chaotic models~\cite{abarbanel2011dynamical,ye2014estimating,rey2014accurate,ye2015improved}, and used to estimate parameters in models of individual biological neurons~\cite{toth2011dynamical,kostuk2012dynamical,meliza2014estimating,kadakia2016nonlinear, breen2016hvc,abarbanel2017unifying}, as well as neutrino interactions in astrophysical settings~\cite{armstrong2017optimization}.  Here we expand beyond applying DA to a single neuron.

The experiments described in this paper are simulations, where \lq\lq data\rq\rq\ are generated from a model whose parameter values are known.  Simulated experiments are a critical first step in testing the applicability of DA to a particular dynamical system, in advance of \lq\lq flying blind\rq\rq\ with real experimental data.  Further, in simulations one may use whichever measurements one desires, regardless of whether such measurements are currently possible to take in a laboratory.  It is in this way that DA lends itself readily to informing experimental design.  In References~\cite{toth2011dynamical,kostuk2012dynamical,meliza2014estimating}, for example, a simulated DA procedure identified which forms of experimentally-injected electrical currents would yield parameter estimations of a desired precision, and the laboratory design was amended accordingly (see \textit{Discussion}).

The model we shall examine for parameter estimation is a small neuronal network that yields a specific pattern of electrical output depending on specific values of electrophysiological and synaptic parameters.  It is a model set forth in Ref~\cite{armstrong2016model}, wherein it corresponded to a functional representation of the avian song-related nucleus HVC.  For the purposes of this paper, the important aspects of the model are twofold.  First, it is biologically relevant in that it exhibits distinct modes of circuit output, depending on parameter values.  Second, it offers a means to quantify the success of the DA procedure: in the predictive phase, the parameter estimates must reproduce the pattern of electrical circuit activity that is associated with the known parameter values.

Specifically, the model is a three-neuron network with all-to-all inhibitory chemical synapses.  DA is employed to estimate 24 parameters: the maximum conductances of the synapses, the synaptic reversal potentials, and the ion channel densities on the membranes of the constituent cells. 

Results are as follows.  The DA procedure identifies two ingredients for successful estimation: 1) the membrane voltages of all three cells are used as measurements, and 2) chaotic current injections are used to stimulate the neurons.  Attempts to use different measurements, including intracellular calcium concentration, are also discussed.  In addition, \lq\lq pruning\rq\rq\ experiments are described, wherein the assumed model has higher dimensionality than that of the model used to generate the provided measurements.  In these cases, DA is able to reduce the assumed model to the appropriate dimensionality.  Finally, we shall discuss the relevance of these simulated experiments to a laboratory setting.

\section{MODEL}

The neuronal network model employed with DA in this paper was chosen for the following feature: when each neuron receives a low-noise (background) current, the circuit engages in one of two specific modes of network activity, where each mode is associated with a particular set of synapse strengths.  These two modes are: i) simultaneous tonic firing of all neurons in the circuit, and ii) a pattern-generating mode wherein the neurons reliably burst in a sequence\footnote{In Ref~\cite{armstrong2016model}, these two modes capture features of the observed population activity in HVC during both quiescence and song, respectively.  For details of the model presented in that paper, see \textit{Appendix 3}.} 

\begin{figure}[H]
  \centering
  \includegraphics[width=86mm]{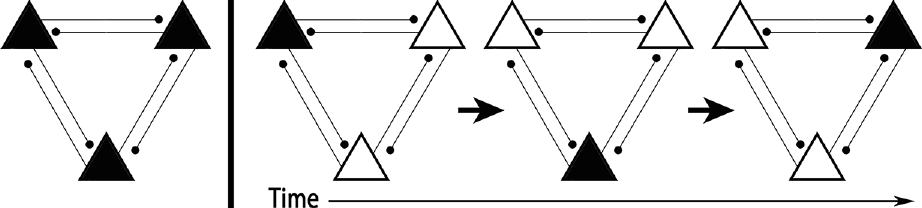}
\caption{{\bf Two functional modes of circuit activity: simultaneous firing and sequential firing}, which can be expressed by the three-neuron structure when it receives a low-amplitude background current.  Each triangle represents one inhibitory neuron, and they are connected all-to-all.  Darkened and white shapes correspond to neurons that are currently active and inactive above spike threshold, respectively.  \textit{Left}: simultaneous firing of the three nodes, for sufficiently low coupling.  \textit{Right}: sequential firing, for a higher range of coupling strengths.}
\label{fig1}
\end{figure}

The model is a three-neuron network with all-to-all inhibitory chemical synapses.  The two modes of electrical activity that will be examined in this paper are shown in Figure~\ref{fig1}.  At left in Figure~\ref{fig1} is depicted synchronous firing of the three neurons, a mode that emerges for sufficiently low synaptic strengths.  At right is depicted a series rotation of firings, due to mutual inhibition that emerges for higher synaptic coupling strengths (for details regarding the mechanism of mutual inhibition, see \textit{Appendix 3}). 

\subsection{\textbf{The neurons}}

The neuron model\footnote{The neuron model has been simplified slightly from its original form in Ref~\cite{armstrong2016model}, for the purposes of this paper.  Two ion channels that are known to exist in these cells have been removed, in order to minimize redundancies while preserving the important feature of these cells that they be capable of both bursting and tonic activity.} is a single-compartment Hodgkin-Huxley-type~\cite{hodgkin1952measurement} construction, based on the electrophysiological studies of HVC inhibitory interneurons~\cite{daou2013electrophysiological} and by ongoing work in the laboratory of Daniel Margoliash at the University of Chicago.  

The time course of membrane voltage for each neuron      
\noindent
\textit{i} is written as: 
\begin{align} 
\label{eq:dVdt}
  C_i\diff{V_{i}(t)}{t} &= I_{L,i}(t) + I_{Na,i}(t) + I_{K,i}(t) + I_{CaT,i} \notag\\&+ \sum_{j \neq i}I_{syn,ij}(t) + I_{inj,i}.
\end{align}
\noindent
The parameter C is the membrane capacitance.  The $I_{syn}$ terms represent synaptic input currents.  $I_{inj}$ is a current injected by the experimenter.  The ion channel currents for the $i^{th}$ neuron are: 
\begin{align} 
\label{eq:ionCurrents}
  I_{L,i}(t) &= g_{L,i}(E_{L,i} - V_i(t))\\ \nonumber
  I_{Na,i}(t) &= g_{Na,i} m_i(t)^3 h_i(t) (E_{Na,i} - V_i(t)) \\\nonumber
  I_{K,i}(t) &= g_{K,i} n_i(t)^4 (E_{K,i} - V_i(t))\\\nonumber
  I_{CaT,i}(t) &= g_{CaT,i} a_i(t)^3 b_i(t)^3 GHK(V_i(t),[Ca]_i(t)),
\end{align}
\noindent
where $GHK(V_i(t),[Ca]_i(t))$ is:
\begin{align*} 
  GHK(V_i(t),[Ca]_i(t)) &= V_i(t)\frac{[Ca]_i(t) - Ca_{ext}e^{-2FV_i(t)/RT}}{e^{-2FV_i(t)/RT} - 1}.
\end{align*}
\noindent
The parameters denoted \textit{g} are the maximum conductances of each current; the parameters denoted \textit{E} are the respective reversal potentials.  [Ca](t) is the intracellular $Ca^{2+}$ concentration as a function of time.  $Ca_{ext}$ is the extracellular concentration of $Ca^{2+}$ ions.  In the GHK current, \textit{F} is the Faraday constant, \textit{R} is the gas constant, and \textit{T} is temperature, which is taken to be 37$^{\circ}$ C.  The gating variables $U_i(t)$ = {m(t), h(t), n(t), a(t), b(t)} satisfy:
\begin{align}
\label{eq:gatingVariables} 
  \diff{U_i(t)}{t} &= (U_{\infty}(V_i(t)) - U_i(t))/\tau_{Ui}(V_i(t))\\\nonumber
  U_{\infty}(V_i) &= 0.5 [1 + \tanh((V_i - \theta_{U,i})/\sigma_{U,i})]\\\nonumber
  \tau_{Ui}(V_i) &= t_{U0} + t_{U1}[1 - \tanh^2((V_i - \theta_{U,i})/\sigma_{U,i})]. 
\end{align}
\noindent
The calcium dynamics evolve as:
\begin{align*} 
  \diff{[Ca_i](t)}{t} &= \phi_i I_{CaT_i} + \frac{Ca_{0,i} - [Ca_i](t)}{\tau_{Ca,i}}.
\end{align*}
\noindent
$Ca_0$ is the equilibrium concentration of calcium inside the cell, and $\phi$ is a constant that summarizes the effects of volume and surface area.  

\subsection{\textbf{The synapses}}

The synapse dynamics follow formalism for chemically-delivered neurotransmitter pulses~\cite{destexhe2001thalamocortical,destexhe1994synthesis}: 
\begin{align} 
\label{eq:synapses} 
  I_{syn,ij} &= g_{ij} s_{ij}(t)(E_{syn,i} - V_{i}(t))\\\nonumber  
  \diff{s_{ij}(t)}{t} &= \nu T(V_j(t))[1 - s_{ij}(t)] - \gamma s_{ij}(t)\\\nonumber   
  T(V_j(t)) &= \frac{T_{max}}{1 + \exp(-(V_{j}(t) - V_{P})/K_{P})}. 
\end{align}  
\noindent
$I_{syn,ij}$ is the current entering cell $i$ from cell $j$\footnote{Ref~\cite{armstrong2016model} adapted the synapse dynamics so that the inhibitory-to-inhibitory connections are functions of a time-varying maximum neurotransmitter concentration $T_{max}$.  In this paper, that detail is omitted.}.  $E_{syn,i}$ is the synaptic reversal potential of cell i, and $s_{ij}$(t) is the synaptic gating variable.  The rate constants $\nu$ and $\gamma$ have units of 1/time.  $V_{P}$ and $K_{P}$ are parameters governing the shape of the distribution of neurotransmitter rise and fall as it drives gating variables $s_{ij}$.  The neurons and synapses are distinguishable via different values of all electrophysiological and kinetic parameters.  For a list of the parameters that were taken to be known and fixed during the D.A. procedure, see \textit{Appendix 4}.

To generate the simulated data, the equations of motion were integrated using Python's odeINT, an adaptive fourth-order Runge-Kutta scheme, using a time step of 0.1 ms.

In this paper, the aim is to infer the six inhibitory maximum conductances $g_{ij}$ that are required to reproduce sequential-bursting mode and simultaneous-firing mode of the circuit.  The synaptic reversal potentials and maximum conductances of ion channels on the three cells are also simultaneously estimated.

\section{DATA ASSIMILATION}

\subsection{\textbf{General formulation}}

Data assimilation is a procedure whereby information in available measurements is used to complete a model of the dynamical system from which the measurements were obtained, where the test of success is the predictive power of the completed model.    The model $\bm{F}$ is a set of \textit{D} ordinary differential equations that evolve in time \textit{t} as:
\begin{align*}
  \diff{x_a(t)}{t} &= F_a(\bm{x}(t),\bm{p}); \hspace{1em} a =1,2,\ldots,D,
\end{align*}
\noindent
where the components $x_a$ of the vector \textbf{x} are the model state variables.  The unknown parameters to be estimated are contained in $\bm{p}$; note that the model evolution depends on $\bm{p}$. 

A subset \textit{L} of the \textit{D} state variables is associated with measured quantities.  One seeks to estimate the $p$ unknown parameters and the time evolution of all state variables during the time window in which the measurements are provided, and to then use those estimates to predict the model evolution at times outside the estimation window.  The prediction phase is the test of estimation quality.

A prerequisite for estimation using real experimental data is the design of simulated experiments, where the true values of parameters are known.  In this stage, one determines whether the design of the DA procedure consistently yields the correct solution.  Note that in addition to providing a consistency check, simulated experiments offer the opportunity to ascertain \textit{which} and \textit{how few} experimental measurements, in principle, are sufficient to complete a model.  

\subsection{\textbf{Optimization framework}}

DA can be formulated as an optimization, where one seeks to find the extremum of a cost function.    We shall take this approach, and write the cost function in two terms: 1) a term representing the difference between state estimate and measurement (measurement error), and 2) a term representing model error.  It will be shown below in this Section that treating the model error as finite offers a systematic method to identify the lowest minimum, in a specific region of state-and-parameter space, of a non-convex cost function.  The surface of the cost function is searched via the variational method.  The procedure in its entirety (that is: a variational approach to minimization coupled with an iterative method to identify a lowest minimum of the cost function) is referred to as variational annealing (VA).  

Applications of VA have included estimating electrophysiological properties of individual neurons in HVC~\cite{meliza2014estimating,kadakia2016nonlinear}, CA1 neurons in a mouse model of Alzheimer's disease~\cite{breen2017use}, and very-large-scale integrated (VLSI) chips whose circuit components were designed as Hodgkin-Huxley neuron models~\cite{wang2016data}.  The procedure has also been used to estimate synapse strengths in a six-neuron circuit with known electrophysiology, but which does not produce a patterned or otherwise stereotyped activity~\cite{knowlton2014}.

The cost function $A_0$ is written as:
\begin{align}
\label{eq:costfunction}
  A_0(\bm{x}(n),\bm{p}) &= \sum_{j=1}^J\sum_{l=1}^{L} \frac{R^{l}_{m}}{2}(y_l(n) - x_{l}(n))^2 \notag\\
&+ \sum_{n=1}^{N-1}\sum_{a=1}^{D} \frac{R^{a}_{f}}{2}\left(x_a(n+1) -
f_a(\bm{x}(n),\bm{p})\right)^2.
\end{align}
\noindent
One seeks the path $\bm{X}^0 = {\bm{x}(0),...,\bm{x}(N),\bm{p}}$ in state space on which $A_0$ attains a minimum value.

The first squared term of Equation~\ref{eq:costfunction} governs the transfer of information from measurements $y_l$ to model states $x_l$.  It derives from the concept of mutual information of probability theory~\cite{abarbanel2013predicting}.  Here, the summation on \textit{j} runs over all discretized timepoints $J$ at which measurements are made, which may be some subset of the all integrated timepoints of the model.  The summation on \textit{l} is taken over all \textit{L} measured quantities.  (For example, one of the simulated experiments of this paper employs as measurements the time series of membrane voltage of all three neurons; for this case, $L=3$.) 

The second squared term of Equation~\ref{eq:costfunction} incorporates the model evolution of all \textit{D} state variables $x_a$\footnote{Incorporating model error, rather than measurement error alone, typically yields a completed model with significantly stronger predictive power (see any general reference on data assimilation, e.g. Refs~\cite{kimura2002numerical,kalnay2003atmospheric,evensen2009data,betts2010practical,whartenby2013number,an2017estimating}).  One can understand the improvement by recognizing that a measurement error considers only the state variables that correspond to measureable quantities.  The model error - by providing a map from measured to unmeasured quantities - requires that the behavior of \textit{all} state variables be consistent with the measurements.}.  (For the network model in this paper, $D=27$: Equations~\ref{eq:dVdt},~\ref{eq:ionCurrents},and~\ref{eq:gatingVariables} describe a seven-dimensional neuron, of which there are three; Equations~\ref{eq:synapses} describe a one-dimensional synapse, of which there are six.)  The second squared term of Equation~\ref{eq:costfunction} can be derived from a consideration of Markov-chain transition probabilities.  The term $f_a(\bm{x}(n))$ is defined, for discretization, as: $\frac{1}{2} [F_a(\bm{x}(n)) + F_a(\bm{x}(n+1))]$.  Here, the outer sum on \textit{n} is taken over all discretized timepoints of the model equations of motion.  The sum on \textit{a} is taken over all \textit{D} state variables.  Note that the second term of Equation~\ref{eq:costfunction} is shorthand.  For the complete formulation, and for a short derivation, see \textit{Appendix 1}.  For a full treatment, see Ref~\cite{abarbanel2013predicting}.

$R_m$ and $R_f$ are inverse covariance matrices for the measurement and model errors, respectively.  In this paper the measurements are taken to be mutually independent\footnote{Of course, for a dynamical system, one would expect measurements taken at successive time steps to be related.  The assumption of independence is taken to simplify the calculation of the cost function, and for many cases it yields a completed model with the predictive power that is desired for the application in question.}.  The simplification , and also the state variables, rendering these matrices diagonal.  For the  purposes of this paper, $R_m$ and $R_f$ are relative weighting terms, their utility will be described immediately below in this Section. 

The procedure searches a $(D \,(N+1)+ p)$-dimensional state space, where \textit{D} is the number of state variables of a model, \textit{N} is the number of discretized steps, and \textit{p} is the number of unknown parameters.  To minimize $A_0$, the variational approach is employed: the first derivative of $A_0$ with respect to the minimizing path is zero, and its second derivative is positive definite.  The procedure is implemented with the open-source Interior-point Optimizer (Ipopt)~\cite{wachter2009short}.  For a link to the user interface with Ipopt that was employed in this paper, see \textit{Appendix 2}, or the github repository: $https://github.com/evearmstrong/OptForNNetwork$.  Ipopt employs a Newton's, or descent-only, search, and a barrier method to impose user-defined bounds that are placed upon the searches.  All simulations were run on a 720-core, 1440-GB, 64-bit CPU Cluster.

The optimization is performed at all locations on a path in the state and parameter space simultaneously\footnote{Thus the procedure imparts no greater importance to a measurement at any particular time over another.}.  It is in this way that information contained in the measurements $y_l$ are transferred to the model term, so as to estimate the vector of unknown parameters $\bm{p}$.  Or: the measurements \lq\lq guide\rq\rq\ the model to the region of its state-and-parameter space in which such measurements are possible.

\subsection{\textbf{Annealing to identify a lowest minimum of the cost function}}

The complete VA procedure involves an iteration that is aimed to identify the set of parameter estimates corresponding to the lowest minimum of the cost function in a user-defined region of state space~\cite{ye2015systematic}\footnote{The lowest minimum is not necessarily the global minimum, as the region of state space being searched is not infinitely large.}.  

\begin{figure}[H]
  \centering
  \includegraphics[width=86mm]{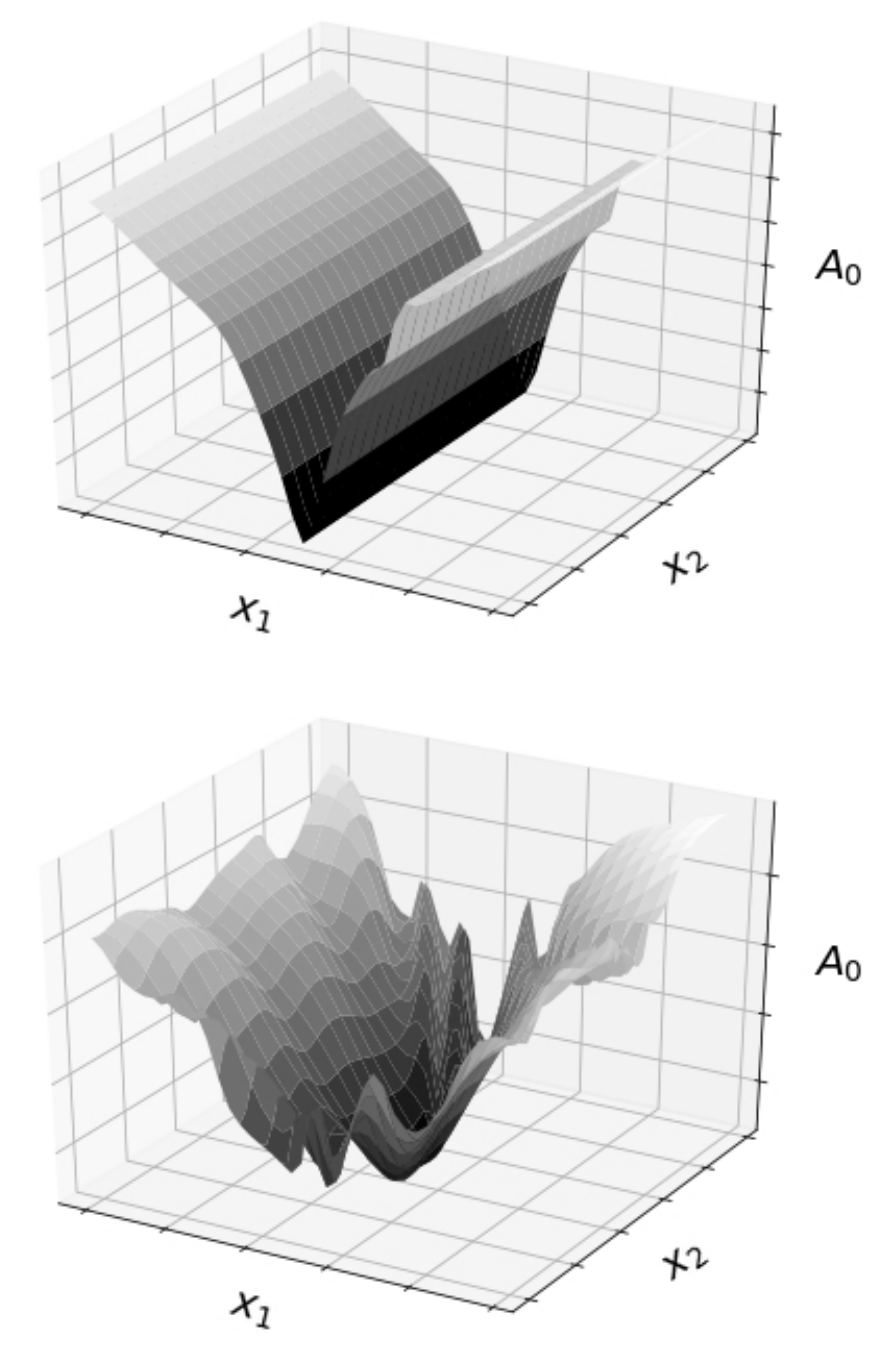}
\caption{{\bf A three-dimensional representation of the cost function ($A_0$) of the form described in Equation~\ref{eq:costfunction} where the model $\bm{f}$ is nonlinear.}  Here the model $\bm{f}$ is a five-dimensional chaotic Lorenz-63 system, which yielded the input currents $I_{inj,i}$ of Equation~\ref{eq:dVdt}.  \textit{Top}: State variable $x_1$ is measured, rendering the surface convex in that direction.  \textit{Bottom}: Neither state variable $x_1$ nor $x_2$ is measured, and $A_0$ is manifestly non-convex.}
\label{fig_action}
\end{figure}

Figure~\ref{fig_action} offers the reader a sense of the difficulty inherent in searching the surface of a cost function of the form described in Equation~\ref{eq:costfunction} where the model $\bm{f}$ is nonlinear and partially observed.  For ease of depiction, this is a three-dimensional representation: the cost function $A_0$ is taken to be a function of two state variables $x_1$ and $x_2$.  The model used to create Figure~\ref{fig_action} is a five-dimensional chaotic Lorenz-63 model, whose outputs were used as the input currents $I_{inj,i}$ of Equation~\ref{eq:dVdt} (see \textit{Simulated Experiments}).  At top in Figure~\ref{fig_action}, state variable $x_1$ is measured, rendering the surface of $A_0$ convex in that direction.  At bottom, neither state variable $x_1$ nor $x_2$ is measured, and multiple solutions exist for each.  For comparison: in this paper the model is 27-dimensional, and - as will be  described below in \textit{Simulated Experiments} - the number of observed state variables is at most three.

One method to minimize the challenge presented by this non-convex search space consists of recursively calculating $A_0$ as the ratio of the model and measurement coefficients is gradually increased.  The mathematical implementation that we choose works as follows. 

We shall first define the coefficient of measurement error $R_m$ to be 1.0, and write the coefficient of model error $R_f$ as: $R_f = R_{f,0}\alpha^{\beta}$, where $R_{f,0} = 0.01$, $\alpha = 1.5$, and $\beta$ is initialized at zero.  $R_f$ is the annealing parameter.  For the case in which $R_f = 0$, relatively free from model constraints the cost function surface is smooth and there exists one minimum of the variational problem that is consistent with the measurements.  We obtain an estimate of that minimum\footnote{The variational approach to seeking the minimum path yields no statistical information about the distribution of cost function levels on paths about that minimizing path $\bm{X}^0$.  In the deterministic limit ($R_f \gg R_m$), however, given adequate measurements $L$, the minimizing saddle paths dominate $A_0$ exponentially~\cite{ye2015systematic}.  Thus, while the problem is formulated statistically, for some procedures the minimizing path yields an excellent approximation without consideration of additional terms.}.  

Then we increase the weight of the model term slightly, via an integer increment in $\beta$.  Beginning a new search at the previously-identified minimum path, we now search a geometry that has been rendered slightly less smooth, via the weak imposition of model dynamics.  We obtain an updated estimate of $A_0$.  We iterate toward the deterministic limit: $R_f \gg R_m$.  Throughout the process, the aim is to remain sufficiently near to the lowest minimum so as not to become trapped in a nearby minimum as the surface of $A_0$ becomes increasingly well-resolved.

This process is employed multiple times in parallel searches, where each search is distinguished by different initial conditions for the state variables and parameters.  The guesses for state variables are randomly selected over the full permitted dynamical range of each variable; the guesses for parameters are drawn randomly from user-defined ranges.  We seek to identify the measurements required for all paths to converge to one solution.  

\subsection{\textbf{Estimation and prediction}}

To perform simulated experiments, the equations of motion are integrated forward to yield simulated data, and the VA procedure is challenged to infer the parameters that were used to generate those data.  Specifically: measurements \textbf{Y} = ${\{\bm{y}(t_0), ... \bm{y}(t_n), ... \bm{y}(t_T)}\}$ are presented to the model at discrete and constant steps between times 0 and T.  The aim is to estimate the state \textbf{X} = $\{{\bm{x}(t_0), ... \bm{x}(t_T)}\}$ and parameters $\bm{p}$ within this estimation window.  

The prediction phase consists of creating a model version in which the true parameters are replaced by the estimates.  The model is integrated forward from the state estimate at the final timepoint of the estimation window.  The resulting time series is compared to the true model evolution, where the true and estimated models receive an identical input current.  Ultimately, the \lq\lq success\rq\rq\ of a prediction must be defined by the aims of a particular investigation.

\section{SIMULATED EXPERIMENTS}

The VA experiments described in this paper were designed to answer the following question: Which measurements of the network, and what forms of stimulating electrical current, are required to yield estimates of connectivity strengths $g_{ij}$, synaptic reversal potentials $E_{rev,i}$, and ion channel maximum conductances $g_{L,i}$, $g_{Na,i}$, $g_{K,i}$, and $g_{CaT,i}$ (where $i\in$[1:3]) that are sufficiently accurate to predict the functional mode of network activity?\footnote{The aim of identifying a metric for successful estimation was also motivated by the fact that the VA method currently provides no error bars on estimates.  A Monte Carlo algorithm is being developed for this purpose~\cite{shirman2018strategic}.}.  All other electrophysiological and kinetic parameters were taken to be known and invariant\footnote{In a laboratory, not all of these parameters will be known to high precision; see \textit{Results E: Estimating additional cellular properties.}} (for these values, see \textit{Appendix 4}).  

The simulated data were generated from two model versions, each defined by a unique set of synapse strengths $g_{ij}$.  The first set corresponds to a mode that - when the injected current $I_{inj}$ is a steady, low-noise back ground current - expresses sequential activations of the interneurons.  These values are on the order of 0.1 $\mu$S (Table~\ref{table1}).  The second set of $g_{ij}$ corresponds to a mode that - with the same steady injected current - expresses simultaneous firing of the neurons.  These values are an order of magnitude lower: roughly 0.01 $\mu$S (Table~\ref{table2}).

For each of the functional modes of circuit activity, multiple versions of the experiment were performed, each version employing a distinct set of measurements.  These measurements were, respectively: 1) membrane voltage of all three neurons; 2) membrane voltage of two out of the three neurons; 3) calcium concentration of all three neurons.  The motivations for these choices will be noted in \textit{Results}.  

In addition, pruning experiments were conducted.  In this procedure, the simulated data are re-generated from a model in which one or more of the unknown parameters - the maximum conductance of either a synaptic connection or an ion channel - is set to zero.  With the model used in the cost function unchanged, the VA procedure is tasked with correctly identifying the appropriate  
\begin{table}[H]
%\begin{table}[!ht]
%\begin{adjustwidth}{-2.25in}{0in} % Comment out/remove adjustwidth environment if table fits in text column.
\small
\centering
\caption{Estimates for $g_{ij}$ corresponding to sequential bursting}
\setlength{\tabcolsep}{6pt}
\begin{tabular}{ l c c c c c c } \toprule
 \textit{Parameter} & Estimated & Correct & Lower & Upper \\
  & value & value & bound & bound \\\midrule   
 \textit{$E_{01}$} & -82.98 & -83.0 & -90.0 & 10.0 \\
 \textit{$E_{02}$} & -83.29 & -83.3 & & \\
 \textit{$E_{10}$} & -82.67 & -82.7 & & \\
 \textit{$E_{12}$} & -82.57 & -82.5 & & \\
 \textit{$E_{20}$} & -83.26 & -83.2 & & \\
 \textit{$E_{21}$} & -82.88 & -82.9 & & \\\hline
 \textit{$g_{01}$} & 0.248 & 0.25 & 0.01 & 10.0 \\
 \textit{$g_{02}$} & 0.403 & 0.4 & & \\
 \textit{$g_{10}$} & 0.283 & 0.28 & &  \\
 \textit{$g_{12}$} & 0.177 & 0.18 & & \\
 \textit{$g_{20}$} & 0.211 & 0.21 & & \\
 \textit{$g_{21}$} & 0.314 & 0.32 & & \\\hline
 \textit{$g_{L,0}$} & 2.88e-3 & 3.0e-3 & 9e-4 & 9e-2 \\
 \textit{$g_{L,1}$} & 3.25e-3 & 3.3e-3 & & \\
 \textit{$g_{L,2}$} & 2.87e-3 & 2.9e-3 & & \\\hline
 \textit{$g_{Na,0}$} & 1.18 & 1.2 & 0.2 & 1.8 \\
 \textit{$g_{Na,1}$} & 0.96 & 1.0 & & \\
 \textit{$g_{Na,2}$} & 1.24 & 1.4 & & \\\hline
 \textit{$g_{K,0}$} & 0.197 & 0.2 & 0.02 & 0.8 \\
 \textit{$g_{K,1}$} & 0.210 & 0.22 & & \\
 \textit{$g_{K,2}$} & 0.150 & 0.17 & & \\\hline
 \textit{$g_{CaT,0}$} & 1.01e-4 & 1.0e-4 & e-5 & e-2  \\
 \textit{$g_{CaT,1}$} & 1.16e-4 & 1.1e-4 & & \\
 \textit{$g_{CaT,2}$} & 1.16e-4 & 9.0e-5 & & \\\bottomrule
\end{tabular}
\newline \small{The columns are: \textit{Estimated value}: parameter estimation from the D.A. procedure; \textit{Correct value}: value used to generate the simulated data that was provided to the D.A. procedure; \textit{Lower bound}: User-imposed lower bound on the parameter value, for the D.A. procedure; \textit{Upper bound}: user-imposed upper bound.  Note that the bounds used for the reversal potentials $E_{ij}$ permit the possibility that synapses are either excitatory or inhibitory.  Units: reversal potentials are in mV; ion channel and synapse maximum conductances are in $\mu$S.  Notation: $g_{ij}$ denotes the weight of the synapse entering cell \textit{i} from cell \textit{j}; $g_{L,i}$ denotes the value of leak current in cell $i$.  Estimates were obtained for annealing parameter values: $R_m = 1$, $R_{f,0} = 0.01$, $\alpha = 1.5$, and $\beta = 27$.  Forty paths were searched, all of which converged to this solution.}\\\hrule
\label{table1}
%\end{adjustwidth}
\end{table}
\noindent
parameters as zero.  This feature of DA estimation has important implications for learning the physics underlying data from a real biological network; see \textit{Discussion}.

For all experiments, the following protocol was used during the estimation window.  The cells received three distinguishable input currents $I_{inj,i}$: the x-, y-, and y-phase-offset output of a chaotic Lorenz-63 model~\cite{lorenz1963deterministic}, with steps spliced into each current at intermittent locations.  The integration time step for the simulated data and the time step of measurement sampling was 0.1 ms; the estimation window was 799 ms.
\begin{table}[H]
%\begin{table}[!ht]
%\begin{adjustwidth}{-2.25in}{0in} % Comment out/remove adjustwidth environment if table fits in text column.
\small
\centering
\caption{Estimates for $g_{ij}$ corresponding to simultaneous firing}
\setlength{\tabcolsep}{6pt}
\begin{tabular}{ l c c c c c c } \toprule
 \textit{Parameter} & Estimated & Correct & Lower & Upper \\
  & value & value & bound & bound \\\midrule   
 \textit{$E_{01}$} & -83.02 & -83.0 & -90.0 & 10.0 \\
 \textit{$E_{02}$} & -83.26 & -83.9 & & \\
 \textit{$E_{10}$} & -82.63 & -82.7 & & \\
 \textit{$E_{12}$} & -82.44 & -82.5 & & \\
 \textit{$E_{20}$} & -83.20 & -83.2 & & \\
 \textit{$E_{21}$} & -82.83 & -82.9 & & \\\hline
 \textit{$g_{01}$} & 0.0256 & 0.025 & 0.01 & 10.0  \\
 \textit{$g_{02}$} & 0.0397 & 0.04 & & \\
 \textit{$g_{10}$} & 0.0281 & 0.028 & & \\
 \textit{$g_{12}$} & 0.0180 & 0.018 & & \\
 \textit{$g_{20}$} & 0.0209 & 0.021 & & \\
 \textit{$g_{21}$} & 0.0321 & 0.032 & & \\\hline
 \textit{$g_{L,0}$} & 2.97e-3 & 3.0e-3 & 9e-4 & 9e-2 \\
 \textit{$g_{L,1}$} & 3.27e-3 & 3.3e-3 & & \\
 \textit{$g_{L,2}$} & 2.90e-3 & 2.9e-3 & & \\\hline
 \textit{$g_{Na,0}$} & 1.18 & 1.2 & 0.2 & 1.8 \\
 \textit{$g_{Na,1}$} & 0.96 & 1.0 & & \\
 \textit{$g_{Na,2}$} & 1.26 & 1.4 & & \\\hline
 \textit{$g_{K,0}$} & 0.196 & 0.2 & 0.02 & 0.8 \\
 \textit{$g_{K,1}$} & 0.212 & 0.22 & & \\
 \textit{$g_{K,2}$} & 0.154 & 0.17 & & \\\hline
 \textit{$g_{CaT,0}$} & 9.90e-5 & 1.0e-4 & e-5 & e-2 \\
 \textit{$g_{CaT,1}$} & 1.08e-4 & 1.1e-4 & & \\
 \textit{$g_{CaT,2}$} & 8.60e-5 & 9.0e-5 & & \\\bottomrule
\end{tabular}
\newline \small{See Caption of Table~\ref{table1} for explanations.}\\\hrule
\label{table2}
%\end{adjustwidth}
\end{table}
During the prediction window, the estimated model was exposed to two novel currents.  The first current was a continuation of the chaotic Lorenz-63 output used in estimation.  The second current was the low-noise step that is known to produce either synchronous or sequential firing, depending on the synapse maximum conductances $g_{ij}$.  For the first, second, and third neuron, this injected background current was: 0.4, 0.5, and 0.3 nA, respectively.

\section{RESULTS}

We shall first examine results using as measurements the voltage traces of all three cells in the network, where the model is the fully-connected network described in \textit{Model}.  For each set of synapse maximum conductances $g_{ij}$, the experiment was performed over forty sets of initial conditions.  All paths converged to a single solution.

Parameter estimates for a model with $g_{ij}$ values yielding i) sequential bursting and ii) simultaneous firing are listed in Tables~\ref{table1} and ~\ref{table2}, respectively.  The parameter estimates were taken at a $\beta$ value of 27, because they resulted in the lowest rms error between simulation and prediction for the measured variables.  (For notes on choosing an optimal ratio $R_f/R_m$, see \textit{Appendix 2}).  The root mean squared (rms) errors for the voltage time series of all three neurons, during both estimation and prediction windows, for all experiments described in this paper are of order $10^1-10^2$ V.  For precise values of rms errors using data generated from the complete model, see the appropriate figure caption.  For precise values of rms errors in pruning experiments, see the appropriate in-text section.

\subsection{\textbf{Result for sequential-bursting mode}}

We shall first discuss estimated and predicted time series of the measured variables for the model in which the synapse maximum conductances $g_{ij}$ were set to the sequential-bursting regime. 
\end{multicols}
\begin{figure}[H]
  \centering
  \includegraphics[width=0.6\textwidth]{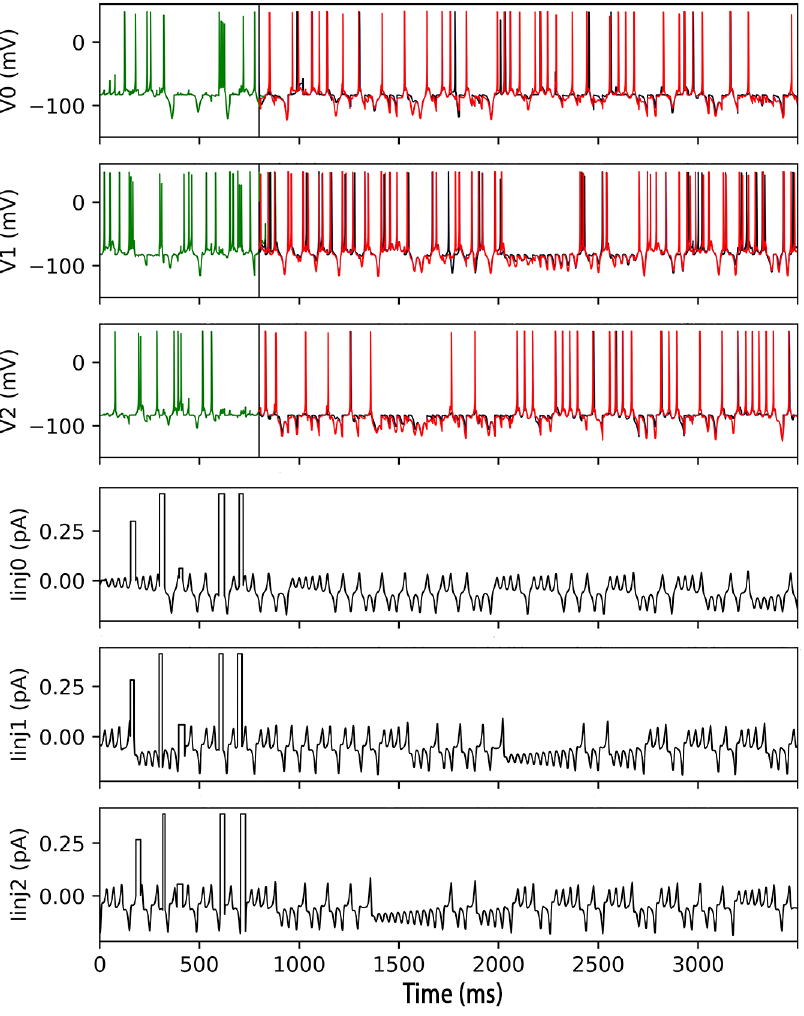}
\caption{{\bf Estimates and predictions of the voltage time series for $g_{ij}$ values corresponding to series activity, where the injected current during the prediction window is a continuation of the chaotic current given during estimation.}  \textit{Top three rows}: Voltage time series of first, second, and third neuron.  Estimate is in green (grey in print); deviation from the true simulation is not visible by eye during the estimation window.  The root mean squared (rms) errors in estimates of the three voltage time series over the estimation window are, respectively, 16.5, 25.2, and 19.5 V.  Prediction begins in red (grey in print) at $t = 799.1$ ms; true simulation is in black. Rms errors during the prediction window are, respectively, 13.7, 23.7, and 14.7 V. \textit{Bottom three rows}: Chaotic injected currents with steps incorporated, given to first, second, and third neuron.  Forty out of forty sets of initial conditions converged to this solution.}
\label{fig2}
\end{figure}
\begin{multicols}{2}
The result using a chaotic current injection in the prediction phase is shown in Figure~\ref{fig2}.   During the estimation window, the true simulation and the estimate are green (light grey in print); that is, indistinguishable by eye (see Caption of Figure~\ref{fig2} for root-mean-squared errors).  The prediction begins in red (light grey in print) at $t = 799.1$ ms, at which time the simulated data are visible in black.  Chaotic current injections do not elicit network activity that is associated with animal behavior, but this result is shown to note that chaotic waveforms for stimulating currents are required to yield estimates that are sufficient for predictive purposes; see \textit{Subsection D: Choosing the injected currents}.

\begin{figure}[H]
  \centering
  \includegraphics[width=0.5\textwidth]{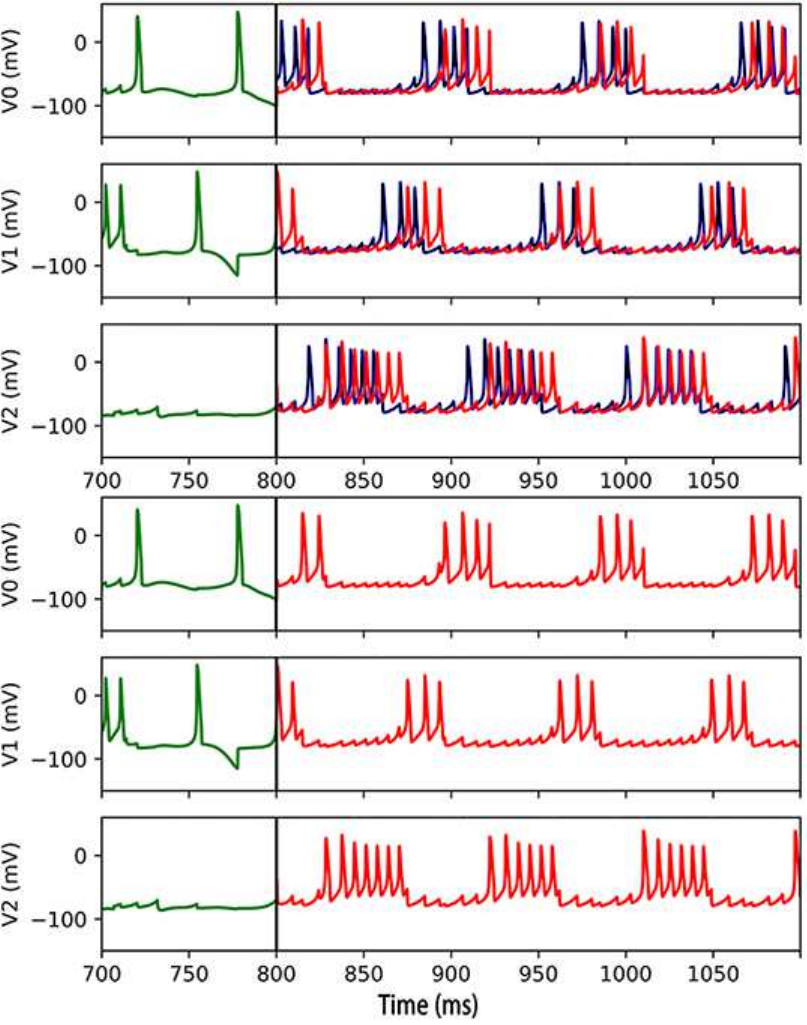}
\caption{{\bf Estimates and predictions of voltage time series for $g_{ij}$ values corresponding to series activity, where the injected current during the prediction window is a step current that yields sequential-bursting activity in the known model.}  Injected currents during prediction, not shown, are: 0.4, 0.5, and 0.3 nA for each neuron, respectively.  \textit{Top three rows}: Prediction (red online and grey in print) is shown along with true simulation (black).  Rms errors for the three voltage traces exposed to these step currents during this prediction window are, respectively, 10.5, 28.1, and 35.5 V - the same order of magnitude as those obtained using the chaotic input current.  \textit{Bottom three rows}: To aid the eye in discerning the series activity, prediction is shown alone.  Forty out of forty sets of initial conditions converged to this solution.} 
\label{fig3}
\end{figure}

More tellingly, for the purposes of this paper: Figure~\ref{fig3} shows results using the steady background current that generates sequential-bursting activity in the known model.  Figure~\ref{fig3} shows two triads of voltage traces.  The top triad shows estimation, prediction, and true model evolution for all three cells.   To aid the eye in identifying the sequential-firing activity, the bottom triad shows the estimation and prediction alone - with true simulation removed.  The neuron order, stability of that order, constancy of the relative phases, spikes per burst for each neuron, and constancy of rotation rate are preserved.  The predicted rotation rate is slightly faster.  This inconsistency may be due either to inherent chaos in the system (in which case even an excellent - but not exact - state estimate may yield a divergent outcome), or to specific parameter estimates that are not exact; see \textit{Discussion}.  

\end{multicols}
\begin{multicols}{2}
\begin{figure}[H]
  \centering
  \includegraphics[width=86mm]{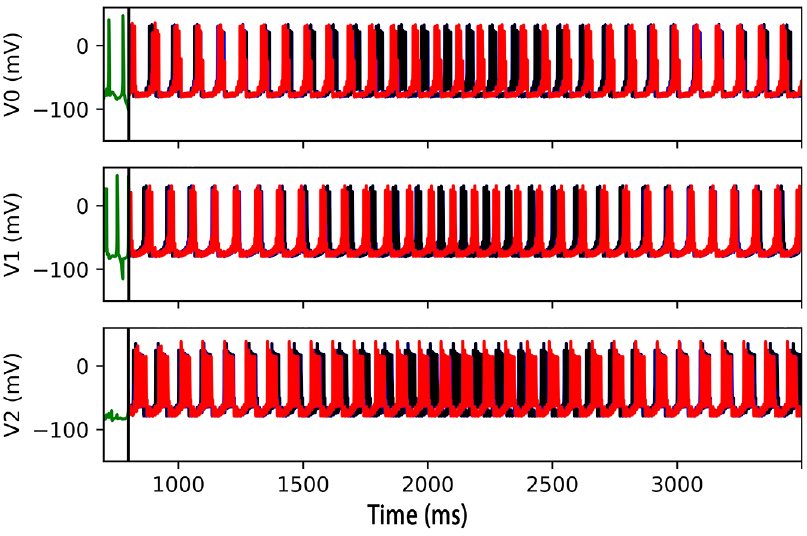}
\caption{{\bf Estimates and predictions of Figure~\ref{fig3}, with an extended x-axis}, showing that the series-firing prediction continues at a stable - and slightly faster - rate of rotation.  Rms errors for the three voltage traces during this extended prediction window are, respectively, 31.0, 28.0, and 34.4 V - of the same order of magnitude as the errors noted in Figure~\ref{fig3}.}
\label{fig4}
\end{figure}
Finally, Figure~\ref{fig4} shows the same result (of Figure~\ref{fig3}) with an extended x-axis.  The series of activations can no longer be discerned; the figure is intended to demonstrate that the activity persists at a reliable rate for 2700 ms\footnote{This duration is roughly ten times longer than the required duration for the associated animal behavior described in Ref~\cite{armstrong2016model}.}.  

\subsection{\textbf{Result for simultaneous-firing mode}}

\begin{figure}[H]
  \centering
  \includegraphics[width=0.6\textwidth]{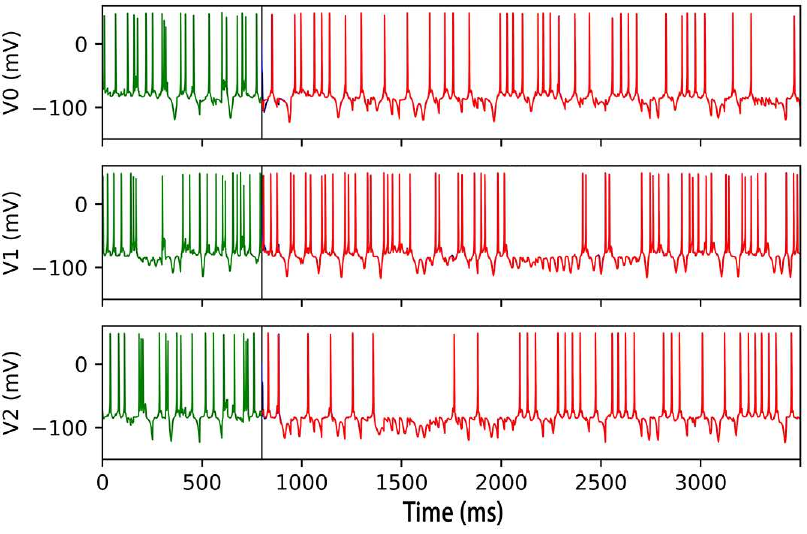}
\caption{{\bf Estimates and predictions of voltage traces for $g_{ij}$ values corresponding to simultaneous firing, where the injected current during the prediction window is a continuation of the chaotic current given during estimation.}  Rms errors for the three voltage time series over the estimation window are, respectively, 13.1, 12.8, and 12.5 V - of the same order of magnitude as the numbers reported in Figure~\ref{fig2} for the sequential-bursting mode.  Forty out of forty sets of initial conditions converged to this solution. }
\label{fig5}
\end{figure}

\begin{figure}[H]
  \centering
  \includegraphics[width=86mm]{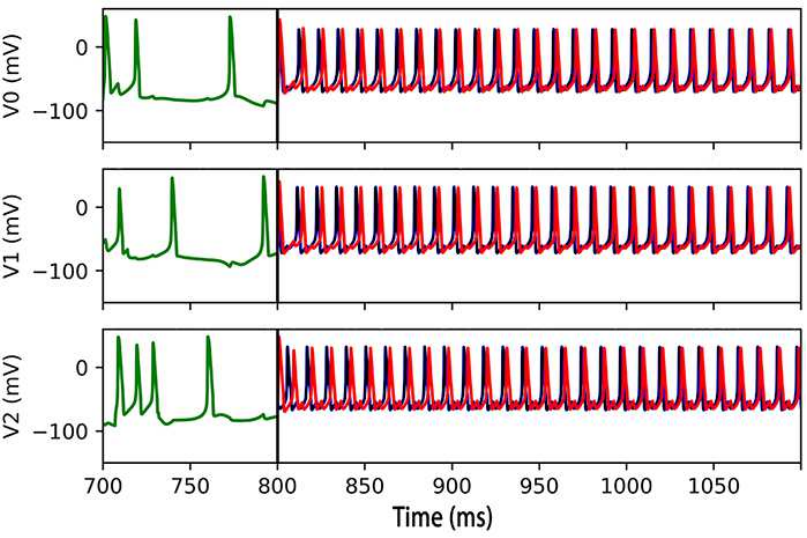}
\caption{{\bf Estimates and predictions of voltage traces for $g_{ij}$ values corresponding to simultaneous firing, where the injected current during the prediction window is the step current that yields simultaneous firing in the known model.}  Rms errors for the three voltage traces during the prediction window are, respectively, 40.5, 39.12, and 39.9 - slightly higher, but of the same order of magnitude, as those reported for the prediction of the sequential-bursting case using a step current (Figure~\ref{fig3}.)  Forty out of forty sets of initial conditions converged to this solution.  This activity is stable over 3500 ms (not shown), with rms errors respectively: 36.0, 35.8, and 36.2 V.} 
\label{fig6}
\end{figure}

Now we shall examine results for the model in which the synapse maximum conductances $g_{ij}$ were set to the simultaneous-firing regime.  Estimation and prediction windows are shown in Figure~\ref{fig5} for the chaotic current and Figure~\ref{fig6} for the constant background.  Figure~\ref{fig6} shows simultaneous firing predicted through a duration of 300 ms; this activity is stable over 3000 ms (not shown).  

\subsection{\textbf{Pruning}}

In a variation of the experiment described above, the simulated data were re-generated using a model in which one or more of the 24 unknown parameters were set to zero.  With the model within the cost function unchanged, the VA procedure was challenged to identify the appropriate parameter values as zero.  This method has been used to determine the existence of ion channel currents in simulated experiments of individual cells~\cite{toth2011dynamical}, as well as the existence of active synaptic connections among cells in a six-neuron network model, given their voltage traces~\cite{knowlton2014}.  This type of study has implications for the experimental observations that the ion channel constituency of neurons can vary considerably across neurons that are
\end{multicols}
\begin{multicols}{2}
\noindent
considered to be of the same \lq\lq type\rq\rq, as can the functional forms of specific ion channel currents (e.g. Ref~\cite{grashow2010compensation}).

Here we shall examine results of pruning the complexity of the network in terms of 1) ion channel constituency, and 2) the network's all-to-all inhibitory connectivity.  For each case, results were consistent over ten trials, and were most accurate for a value of 27 for the annealing parameter $\beta$.  In both cases, estimates were of an accuracy similar to that of the original experiment.

\subsubsection{Pruning ion channels}

As a test of the VA procedure's ability to identify an unnecessarily complex neuron model, the maximum conductance of the \textit{CaT} current in \lq\lq Neuron 0\rq\rq\ (of 0, 1, and 2) was set to zero.  All other parameter values were unchanged, and the three voltage traces were given as measurements.  The experiment was performed once using the synapse strengths associated with sequential-bursting mode, and once using those associated with simultaneous-firing mode.  As with the original experiments, the estimation window employed chaotic input currents over an 8000-point time series with a step size of 0.1 ms.  

Estimates of the maximum conductances $g_{CaT}$ of the three neurons are shown in Table~\ref{table3}.  For estimates of all parameters, see \textit{Appendix 4}.  Estimates of $g_{CaT}$ for Neuron 0 were $5.86 \times 10^{-7}$ and $8.26 \times 10^{-8}$ for each mode, respectively: roughly four orders of magnitude lower than the maximum conductances of $g_{CaT}$ for the other two cells.  The rms errors in estimates of the three voltage traces for the sequential-bursting and simultaneous-firing regimes were, respectively: 5.4, 10.8, and 9.6 V, and 9.5, 26.6, and 20.0 V - of the same order of magnitude as the errors reported for the previous experiments described in this paper.

\subsubsection{Pruning synaptic connectivity}

Next it was examined whether a sparsely-connected network would yield estimates of the accuracy found in the original experimental design.  To this end, four synapse maximum conductances were set to zero.  These were: $g_{01}$, $g_{02}$, $g_{10}$, and $g_{12}$.  As with the pruning of the CaT ion channel, this experiment was performed once using sequential-bursting- and once using simultaneous-firing-mode synapse strengths, all other parameter values were unchanged, and the measurements provided were the three voltage traces.  

Results were similar to the case for $g_{CaT} = 0$.  Estimates for the $g_{ij}$ of the six inhibitory connections are shown in Table~\ref{table4}.   The parameter values that have been changed from the original design are underlined.  For estimates of all parameters, see \textit{Appendix 4}.  The rms errors in estimates of the three voltage traces for the sequential-bursting and simultaneous-firing regimes during the estimation window were, respectively: 2.3, 6.3, and 8.83 V, and 8.6, 6.4, and 2.9 V.  The procedure was also repeated with each individual conductance $g_{ij}$ set to zero in the generative model; estimates were of similar accuracy (not shown).
\end{multicols}
\begin{table}[H]
%\begin{adjustwidth}{-2.25in}{0in} % Comment out/remove adjustwidth environment if table fits in text column.
\small
\centering
\caption{Estimates of maximum conductances $g_{CaT}$, with $g_{CaT,0}$ set to zero}
\setlength{\tabcolsep}{6pt}
\begin{tabular}{ l c|c|c c c } \toprule
 \textit{Parameter} & Estimate using & Estimate using & Correct & Lower & Upper \\
  & high $g_{ij}$ & low $g_{ij}$ &  & bound & bound \\\midrule   
 \textbf{$g_{CaT,0}$} & 5.86e-7 & 8.26e-8 & \underline{0.0} & 0.0 & e-2  \\
 \textit{$g_{CaT,1}$} & 1.17e-4 & 1.09e-4 & 1.1e-4 & & \\
 \textit{$g_{CaT,2}$} & 1.01e-4 & 9.14e-5 & 9.0e-5 & & \\\bottomrule
\end{tabular}
\newline \small{Estimates of maximum conductances $g_{CaT}$, with $g_{CaT,0}$ set to zero - once using the $g_{ij}$ underlying sequential bursting, and once using those for simultaneous firing.  For estimates of all parameter values, see \textit{Appendix 4}.}\\\hrule
\label{table3}
%\end{adjustwidth}
\end{table}
\begin{table}[H]
%\begin{adjustwidth}{-2.25in}{0in} % Comment out/remove adjustwidth environment if table fits in text column.
\small
\centering
\caption{Estimates of maximum conductances $g_{ij}$, with four $g_{ij}$ set to zero}
\setlength{\tabcolsep}{6pt}
\begin{tabular}{ l c c|c c|c c} \toprule
 \textit{Parameter} & \multicolumn{2}{c|}{Model with high $g_{ij}$ values} & \multicolumn{2}{c|}{Model with low $g_{ij}$ values} & Lower & Upper \\
  & \textit{estimate} & \textit{true} & \textit{estimate} & \textit{true} & bound & bound \\\midrule   
 \textbf{$g_{01}$} & 1.0e-4 & \underline{0.0} & 1.4e-3 & \underline{0.0} & 0.0 & 10.0 \\
 \textit{$g_{02}$} & 0.0 & \underline{0.0} & 3.7e-5 & \underline{0.0} & & \\
 \textit{$g_{10}$} & 8.84e-6 & \underline{0.0} & 4.8e-6 & \underline{0.0} & & \\
 \textit{$g_{12}$} & 0.0 & \underline{0.0} & 0.0 & \underline{0.0} & & \\
 \textit{$g_{20}$} & 0.193 & 0.21 & 0.021 & 0.021 & &\\ 
 \textit{$g_{21}$} & 0.350 & 0.32 & 0.032 & 0.032 & & \\\bottomrule
\end{tabular}
\newline \small{Estimates of synapse maximum conductances $g_{ij}$, with four $g_{ij}$ set to zero - once using the $g_{ij}$ underlying sequential bursting, and once using those for simultaneous firing.  For estimates of all parameter values, see \textit{Appendix 4}.}\\\hrule
\label{table4}
%\end{adjustwidth}
\end{table}
\begin{multicols}{2}
\subsection{\textbf{Choosing the injected currents}}

For neuronal models, the accuracy of state and parameter estimates is sensitive to the choice of stimulating current ($I_{inj,i}$ in Equation~\ref{eq:dVdt})  (e.g.~\cite{toth2011dynamical,kostuk2012dynamical,kadakia2016nonlinear}).  The injected currents for parameter estimations that are shown in Figures 2-6 were chosen with the following three considerations in mind\footnote{Note that these \lq\lq required\rq\rq\ traits for an effective stimulating current have not been systematically studied; rather the citations above refer to informal comments regarding results.  To perform the study rigorously, one should consider the mutual information between stimulus and response, and between stimulus and prediction, across a range of stimuli.}.  

First, the modulation must be slower than the intrinsic dynamics of the neuron.  Otherwise, the neuron will behave as a low-pass filter, and the experimenter will not know precisely what current the neuron \lq\lq sees\rq\rq.

Second, to avoid problems associated with symmetries in the model equations of motion, each neuron should receive a distinct current.

Third, the current's amplitude and frequency must vary sufficiently so as to force the model's degrees of freedom to explore their full dynamical ranges.  This was in fact the motivation to add the intermittent steps to the chaotic currents used in this paper.  Namely: the model neurons are known to be capable of bursting (as well as spiking), but the chaotic currents alone did not produce bursts.  The steps do produce bursts, and thus exposing the neuron to a combination of chaotic and step forms more faithfully demonstrates the variety of electrical output of which the dynamics are capable.  Once the steps were added, estimates of the ion channel maximum conductances improved by roughly a factor of two.  By contrast, the use of step or sinusoidal currents failed to produce converging solutions.

\subsection{\textbf{Estimating additional cellular properties}}

Still using as measurements the three voltage time series, and resetting all parameters to their original (non-zero) values for sequential-bursting mode, the list of parameters to be estimated was extended to include the time constants $t_{U0}$ and $t_{U1}$ of all gating variables of the model (refer to \textit{Model} for the forms of the gating variables).  With five gating variables, the additional number of unknown parameters is ten (see Equations~\ref{eq:ionCurrents} and~\ref{eq:gatingVariables}).  This addition resulted in high degeneracy of estimates and poor predictive power across trials (see \textit{Discussion}). 

\subsection{\textbf{Identifying the measurements required for successful estimation}}

Keeping as measurements the membrane voltages, a variation on the experiment was performed in order to ascertain whether a successful estimation requires the sampling of all neurons within the network.  This assumption originates from a well-known problem in causality inference that all components of a system must be sampled in order to recover complete information about the system\footnote{In causal inference, any causal link $A \rightarrow B$ can be replaced with $A \rightarrow Z \rightarrow B$ while preserving the causal effect.  Measuring A alone is insufficient; one must also measure all possible Z (e.g. Ref~\cite{pearl2000causality}).}.  To this end, the membrane voltage time series of just two out of the three neurons were provided as measurements.  

This design was indeed inadequate for prediction\footnote{The ion channel maximum conductances and reversal potentials of the \textit{measured} cells, and the $g_{ij}$ of synapses \textit{leaving} the measured cells were estimated to an accuracy comparable to the estimates of Tables~\ref{table1} and ~\ref{table2}.  The corresponding values for the \textit{unmeasured} cell were poor, and predictions of voltage time series for all three cells were poor (not shown).}.  The implication for experiment is that, in order to obtain parameter estimates sufficiently accurate to predict network electrical output, all cells of the circuit must be sampled during the estimation procedure.  It is fortuitous, then, that there exist some isolatable functional circuits in crustaceans whose constituent cells are sufficiently large in volume - and sufficiently small in number - to facilitate simultaneous whole-cell recordings of all cells within the circuit; see \textit{Discussion}.

Finally, given the wide use of calcium imaging in neuroscience, a variation of the original experiment was performed to ascertain whether intracellular calcium concentration alone contained sufficient information to yield adequate predictions.  That is, the measurements used were the time series of calcium concentration for all three cells.  

Results were poor.  Over forty trials tested, no two paths converged to a common solution, and many purported \lq\lq estimates\rq\rq\ corresponded to the user-imposed search bounds.  Further, the iterative procedure of increasing model error coefficient $R_f$ was extremely slow compared to the experiments in which voltage traces were used as measurements.  For 36 out of 40 trials, the procedure did not complete within the 72-hour allotment permitted when using shared computing resources.  By comparison, the experiments using voltage traces required roughly 20 hours.  The implication is that the transfer of information from measurements to the unmeasured model state variables is vastly less efficient for the case of calcium concentration than for voltage trace.  This finding is consistent with the model equations of motion.  While membrane voltage (Equation~\ref{eq:dVdt}) is a direct function of the 24 parameters to be estimated, the calcium concentration relates directly only to the CaT-type ion channel maximum conductance. 

\section{DISCUSSION}

The simulated experiments presented in this paper indicate that data assimilation provides a recipe for identifying both the measurements and forms of stimulating current that are necessary to simultaneously estimate several tens of parameters governing a model biological neuronal network, if the test of estimate quality is the ability of the estimated model to predict a particular mode of circuit activity.  Further, DA has the potential to reduce the dimensionality of an assumed model, given the information contained in measurements.  The procedure did encounter difficulty when the number of unknown parameters was increased by ten.  In this Section, we shall address that difficulty, and consider the applicability of DA in a laboratory.

\subsection{\textbf{Information for degeneracy breaking}}

As noted, when the list of parameters was extended to include time constants of ion channel gating variables, multiple sets of estimates were obtained.  This result indicates that insufficient information existed in the provided measurements for breaking degeneracies associated with multiple minima of the cost function.  There exist several possible remedies for this problem.

One means to provide additional information to the DA procedure is the use of more than one \lq\lq training pair\rq\rq.  In the experiments described in this paper, one pair was used: a set of three injected currents, and a set of three voltage traces (input and output, respectively).  Supplying the network model with a collection of such pairs, where various injected currents capture a range of waveforms, frequencies, amplitudes, and temporal durations, may better-resolve the surface of the cost function. 

A second means may be the use of time delays.  Time delay embedding is a method to extract additional information from existing measurements, by examining the relationship between successive samples of each measurement (see Ref~\cite{rey2014accurate} for an exercise in using time delays within a DA procedure similar to the form used in this paper).  

Third, we have assumed that in the deterministic limit ($R_f \gg R_m$), the minimizing saddle paths dominate $A_0$ exponentially.  Indeed, for some procedures the minimizing path yields an excellent approximation of the expectation value of a path without the consideration of additional terms~\cite{ye2015systematic}.  This is not guaranteed to be the case.  It is possible that the inclusion of additional terms in the calculation of the cost function would mitigate degeneracies. 

\subsection{\textbf{Implications for experiment}}

Let us now extend our discussion of data assimilation into the laboratory setting.  For the case of individual neurons, DA has been found to inform experimental design.  In References~\cite{toth2011dynamical,kostuk2012dynamical,meliza2014estimating}, those authors identified which forms of injected current yielded estimates of electrophysiological parameters - including ion channel maximum conductances and gating variable time constants - of a desired precision.  Using simulated data, they found that currents with chaotic waveforms significantly outperformed step and ramp currents, and the laboratory procedure was adjusted accordingly.  Now the applicability of DA to the study of small circuits is considered.  

As noted, for roughly three decades, small-circuit studies have been key to unveiling fundamental electrophysiological and synaptic processes in neuroscience, and how they give rise to patterned electrical activity underlying rhythmic motor behavior.  Some such circuits are identified as central pattern generators (CPGs), in that they are capable of engaging in self-sustained sequenced activity for a finite time, without the need for continued external stimulation or sensory information.  This capability appears to be a species-invariant property of the central nervous system, and one that has evolved independently multiple times in the animal kingdom~\cite{delcomyn1980neural,ijspeert2008central}.  

Some invertebrate CPGs possess the following properties: 1) they are comprised of fewer than $\sim$ ten neurons that can be repeatedly identified across animals, 2) the neurons are sufficiently large that simultaneous intracellular recordings may be obtained from each, 3) the neurons are localized within a small anatomical area, facilitating their complete isolation from the animal, and 4) most of the neurons are also motor neurons, and thus the network activity may be directly correlated with behavior.  Examples include a detailed exploration of structural connectivity in a six-neuron CPG within the stomatogastric ganglion (STG) of the spiny lobster \textit{Pannulirus interruptus}~\cite{mulloney1974organization}, a four-neuron CPG underlying swimming in \textit{Dendronotus iris}~\cite{sakurai2016central}, and all work previously cited in this paper by the laboratory of Eve Marder on STG of the crustacean \textit{Cancer borealis}.  Such systems may offer an opportunity to take an elaboration of the procedure described in this paper into a laboratory\footnote{For those interested in applying the technique described in this paper to large circuits (i.e. those comprised of tens of neurons): this computational task quickly becomes prohibitively expensive.  In the formulation presented in this paper, we seek a high degree of physical realism, namely in the form of high-dimensional neurons whose dimensions directly correspond to membrane properties.  For a D-dimensional model comprised of $N$ neurons, each new neuron added will contribute its neuronal dimensions, and - for a fully-connected model - 2$\times N$ new reciprocal synapse dimensions.  A more tractable line of attack for such circuits would be a method of model reduction.}

Ref~\cite{marder2016complicating}, for example, describes three simultaneous intracellular voltage traces obtained from cells in the STG of \textit{Cancer borealis} - data that are qualitatively similar to the simulated measurements used in this paper.  Those authors found that various underlying functional connectivity schematics were able to explain observed voltage traces.  It would be worthwhile to ascertain whether DA identifies the correct circuitry, which those authors later identified uniquely via photoinactivation following the recordings\footnote{The results in this paper in fact show that small differences in parameter values can result in similar - but not \textit{identical} - patterned activity.  For the sequential-firing-mode of the circuit, the parameter estimates (Table~\ref{table1}) accurately predicted competitive activity among the neurons, but there was a minor difference in rotation rate (Figure~\ref{fig3}).  It would be interesting to examine the effects - if any - of such differences in electrical circuit output on motor behavior.}.

\subsubsection{\textbf{Examining redundancy}}

The last paragraph above concerns the redundancy inherent in the nervous system.  Across animals, there exists large variability in synaptic and electrophysiological properties while behavioral output is maintained~\cite{calabrese2011coping, marder2006variability}.  Within an animal, redundancy has been described as \lq\lq compensation\rq\rq, where one parameter readjusts in response to a change in another, so that network activity is preserved~\cite{grashow2009reliable,grashow2010compensation,marder2016complicating}.  Further, there exists evidence that reduncancy exists not only in parameter values, but in the form of the model.  For example, the ion channel constituency of neurons can vary considerably across neurons that are considered to be of the same functional \lq\lq type\rq\rq, as can the functional forms of specific ion channel currents (e.g. Ref~\cite{grashow2010compensation}).  If redundancy is aimed to preserve behavior, one may ask: do there exist \lq\lq rules\rq\rq\ governing redundancy?

A DA experiment aimed to tackle this question might run as follows.  A single DA estimation experiment is performed multiple times using different sets of measurements, each set from a distinct animal.  One then compares the estimates - including any model pruning that occurs - across cases.  In doing so, one might consider reformulating the parameter estimation question in terms of \textit{sets of} parameters, rather than a single \lq\lq correct\rq\rq\ set, and seek possible rules governing all sets.  There exists a wealth of work on identifying optimal parameterized model families, rather than one \lq\lq best\rq\rq\ model (see Refs~\cite{balasubramanian1997statistical} and \cite{bialek2001complexity}, and references therein)\footnote{In producing the results presented in this paper, an attempt was made to quantify the sensitivity of the model to particular parameter values, via the variation of one parameter at a time, and then two simultaneously.  The futility of this effort soon became evident.  Ref~\cite{rotstein2016dynamic} approached this problem via a phase space analysis where two pairs of cellular properties were examined at a time.  They found that balances between particular pairs were required to maintain a stable \lq\lq duty cycle\rq\rq.  The expansion of this technique to many dimensions might offer insight into the results of DA estimates across multiple data sets.}.

%\subsubsection{\textbf{Assumptions of model dimensionality}}

%We have examined the pruning of model dimensionality.  Note that inference procedures do not provide the means to do the reverse: to infer that an assumed model contains insufficient complexity to describe the associated data.  Machine learning techniques do exist to prune \lq\lq infinite\rq\rq-dimensional models~\cite{hinton2015distilling}, but - in addition to the computational expense - such complex models are not likely to offer biological insight.  Alternatively, there exist methods to determine the maximum dimensionality contained in a data set, for example: phase space reconstruction~\cite{abarbanel2012analysis}\footnote{Phase space reconstruction is distinct from linear methods such as Principal Component Analysis, in that one preserves any nonlinearities present in the model, counts the complete number of dimensions (rather than employing an arbitrary cut-off), and does not rank the dimensions in importance.}.  Such a procedure might be found to complement a DA procedure in cases where dimensionality is a looming question.  One must take care, however, in defining the aim and scope of calculating dimension based on measurements, as the calculation assumes that the measurements represent the full dynamical range of the underlying model.

\section{ACKNOWLEDGMENTS}
Thank you to Henry Abarbanel, Paul Rozdeba, and Sasha Shirman for discussions on data assimilation and its connection to machine learning.  Thanks to Arij Daou, Joshua Gold, Jorge Golowasch, Michael Long, Eve Marder, Daniel Margoliash, Richard Mooney, and Michael Nusbaum for information on experimental capabilities and thought-provoking discussions. 

\section{Appendix 1: Derivation of the cost function used for data assimilation}
\label{appendix1}

\subsection{\textbf{Purpose and strategy}}

Here we lay out a derivation of the cost function $A_0$ used in this paper.  For a thorough treatment, see Ref~\cite{abarbanel2013predicting}.

We begin by seeking the probability of obtaining a path $\bm{X}$ in the model's state space given observations $\bm{Y}$, or: $P(\bm{X}|\bm{Y})$.  Writing:
\begin{equation*}
  P(\bm{X}|\bm{Y}) = e^{-A_0(\bm{X},\bm{Y})},
\end{equation*}
we mean: \textit{the path $\bm{X}$ for which the probability (given $\bm{Y}$) is greatest is the path that minimizes $A_0$}.  Now, if $A_0$ is sufficiently large (where \lq\lq sufficiently\rq\rq\ must be defined by the results of a particular D.A. experiment using a particular model), we can use Laplace's method to estimate the minimizing path on the surface of $A_0$.  Laplace’s method was developed to approximate integrals of the form: $\int e^{Mf(x)}dx$.  For sufficiently high values of the coefficient $M$, significant contributions to the integral will come only from points in a neighborhood around the minimum, which can then be estimated.    

A formulation for $A_0$ will permit us to obtain the expectation value of any function $G(\bm{X})$ on a path $\bm{X}$.  Expectation values are the quantities of interest when the problem is statistical in nature.  We can write the expectation value of $G(\bm{X})$ as:
\begin{align}
  \langle G(\bm{X}) \rangle &= \frac{\int d\bm{X} G(\bm{X}) e^{-A_0(\bm{X},\bm{Y})}}{\int d\bm{X} e^{-A_0(\bm{X},\bm{Y})}}.
\end{align}
\noindent
That is: the expectation value can be expressed as a weighted sum over all possible paths, where the weights are exponentially sensitive to $A_0$.  The RMS variation, and higher moments of $G(\bm{X})$, can be calculated by taking the $x_a$ to the appropriate higher exponents.  If the quantity of interest is the path $\bm{X}$ itself, then we choose $G(\bm{X}) = \bm{X}$.  

It remains, then, to write a functional form for $A_0$.  This will take place in two steps.  First we shall consider how measurements and model dynamics enter into the process state and parameter estimation.  This we will do via an examination of Bayesian probability theory and Markov chain transition probabilities, for the effect of measurements and model dynamics, respectively.  Second, we shall make four simplifying assumptions: 1) the measurements taken at different times are independent; 2) both measurement and model errors have Gaussian distributions; 3) each measurement is taken to correspond directly to one model state variable; 4) the minimizing path is independent of the guess - in state and parameter space - of the initial path.  

In what follows, we shall describe this strategy.  To remind the reader of the notation: The model consists of $D$ PDEs, each of which represents the evolution of one of the model's $D$ state variables.  From the corresponding physical system, we are able to measure $L$ quantities, each of which corresponds to one of the model's $D$ state variables.  Typically the measurements are sparse ($L \ll D$), and the sampling may be infrequent or irregular.  

\subsection{\textbf{Considering model dynamics only (no measurements yet)}}

We shall first examine this formulation by considering the model's time evolution in the absence of measurements.  We represent the model's path through state space as the set $\bm{X} = \{\bm{x}(t_0),\bm{x}(t_1),\ldots,\bm{x}(t_N),\bm{p}\}$, where $t_N$ is the final \lq\lq time point\rq\rq\ and the vector $\bm{x}(t)$ contains the values of the $D$ total state variables, and $\bm{p}$ are the unknown parameters (here, the phrasing \lq\lq time\rq\rq\ can also be taken to represent other grid parameterizations; for instance: location).

\subsubsection{Assuming that a Markov process underlies the dynamics}

If we assume that the dynamics are memory-less, or Markov, then $\bm{x}(t)$ is completely determined by $\bm{x}(t-\Delta t)$, where $t-\Delta t$ means: \lq\lq the time immediately preceding $t$\rq\rq\ and an appropriate discretization of time $\Delta t$ for our particular model has been chosen.  A Markov process can be described in the continuous case by a differential equation, or as a set of differential equations:
\begin{align*}
  \diff{x_a(t)}{t} &= F_a(\bm{x}(t),\bm{p}); \hspace{1em} a =1,2,\ldots,D,
\end{align*}
and we note that the model is an explicit function of the state variables $\bm{x}(t)$ \textit{and the unknown parameters $\bm{p}$}.  It is in this way that the unknown parameters are considered to be on equal footing with the variables; namely: they are variables with trivial dynamics.

In discrete time, that relation can be written in various forms.  For our purposes, we use the trapezoidal rule:
\begin{align*}
  x_a(n+1) = x_a(n) + \frac{\Delta t}{2}[F_a(\bm{x}(n+1)) +  F_a(\bm{x}(n))],
\end{align*}
where for simplicity we have taken $n$ and $n+1$ to represent the values of $t_n$ and $t_{n+1}$.%Note that by using this formulation, we are treating the LaGrangian as a function of the values of \bm{x} at two timepoints, rather than as a function of \bm{x} and its first derivative $\dot{x}$.

\subsubsection{Permitting stochasticity in the model and recasting its evolution in terms of probabilities}

We are interested in ascertaining the model evolution from time step to time step, where now we allow for some stochasticity in the model dynamics. In this scenario, the evolution can be formulated in terms of transition probabilities, for example: $P(\bm{x}(n+1)|\bm{x}(n))$---the probability of the system reaching a particular state at time $n+1$ given its state at time $n$.  If the process were deterministic, then in our case $P(\bm{x}(n+1)|\bm{x}(n))$ would simply reduce to: $\delta^D (\bm{x}(n+1) - \bm{x}(n) - \frac{\Delta t}{2}\left[\bm{F}(\bm{x}(n+1)) + \bm{F}(\bm{x}(n))\right])$.  We will revisit to this expression later in this Appendix.

For a Markov process, the transition probability from state $\bm{x}(n)$ to state $\bm{x}(n+1)$ represents the probability of reaching state $\bm{x}(n+1)$ given $\bm{x}(n)$ and $\bm{x}$ at \textit{all} prior timesteps.  Or:
\begin{align*}
  P(\bm{x}(n+1)|\bm{x}(n)) &= P(\bm{x}(n+1)|\bm{x}(n),\bm{x}(n-1),\ldots,\bm{x}(0))
\end{align*}
so that
\begin{align*}
  P(\bm{X}) &\equiv P(\bm{x}(0), \bm{x}(1),\ldots, \bm{x}(N)) \\
			&= \prod_{n=0}^{N-1} P(\bm{x}(n+1)|\bm{x}(n)) P(\bm{x}(0))
%  &= \Pi_{n=0}^{N-1}P(\bm{x}(n+1)|\bm{x}(n)) P(\bm{x}(0)).
\end{align*}.

We now write 
\begin{align*} 
  P(\bm{X}) \equiv e^{-A_0(\bm{X})}, 
\end{align*}
\noindent
where $A_0$ is the cost function defined on the model's path $\bm{X}$ in state space.  Or: \textit{the path that minimizes the cost function is the path most likely to occur}.  (The reader might find it of interest to note the quantum-mechanical analog of the transition probability, which involves the trivial addition of the term $\frac{i}{\hbar}$ in the exponent: $P(\bm{x}(n+1)|\bm{x}(n)) = e^{\frac{i}{\hbar}A(t_{n+1},t_n)}$, where $A$ here is the classical action).  Then the model term of the cost function, $A_{0,\text{model}}$, can be written:
\begin{align*}
  A_{0,\text{model}} = -\sum \log[P(\bm{x}(n+1)|\bm{x}(n))] - \log[P(\bm{x}(0))],
\end{align*}
\noindent
where the second term represents uncertainty in initial conditions.

\subsection{\textbf{Now with measurements}}

We now consider the effect of measurements.  Let us define a complete set of measurements $\bm{Y}$ to be the set of all vectors $\bm{y}(n)$ at all times $n$---the analog of $\bm{X}$ for the complete set of state variable values.  We shall examine the effect of these measurements upon a model's dynamics by considering conditional mutual information (CMI).  The reader may find an intuitive understanding of our use of the CMI by the following consideration.  The overall information, in bits, in a set $A$ is defined as the Shannon entropy $H(A) = -\sum_A P(A) \log[P(A)]$.  The CMI is a means to quantify the amount of information, in bits, that is transferred along a model trajectory within a particular temporal window.  That information is equivalent to: $-\sum_{n=0}^N \log[P(\bm{x}(n)|\bm{y}(n),\bm{Y}(n-1))]$.  

The expression $\text{CMI}(\bm{x}(n),\bm{y}(n)|\bm{Y}(n-1))$ asks: \lq\lq How much is learned about event $\bm{x}(n)$  upon observing event $\bm{y}(n)$, conditioned on having previously observed event(s) $\bm{Y}(n-1)$?\rq\rq.  The CMI can be quantified as:
\begin{align*}
\text{CMI}(&\bm{x}(n),\bm{y}(n)|\bm{Y}(n-1)) \\
&= \log\left[\frac{P(\bm{x}(n),\bm{y}(n)|\bm{Y}(n-1))}{P(\bm{x}(n)|\bm{Y}(n-1)) P(\bm{y}(n)|\bm{Y}(n-1))}\right].
\end{align*}

\subsection{\textbf{The complete cost function}}

With measurement considerations included, the cost function now becomes: 
\begin{align*}
  A_0(\bm{X},\bm{Y}) = -\sum \log[P(\bm{x}(n+1)|\bm{x}(n))] - \log[P(\bm{x}(0))] \\
  - \sum \text{CMI}(\bm{x}(n),\bm{y}(n)|\bm{Y}(n-1)),
\end{align*}
\noindent
where the first and second terms represent the model dynamics including initial conditions, and the third term represents the transfer of information from measurements.  The summations are over time.  As noted, this formulation positions us to calculate the expectation value of any function $G(\bm{X})$ on the path $\bm{X}$.  

We now offer an interpretation of the measurement term.  The measurement term can be considered to be a nudging (or synchronization) term.  While nudging terms are often introduced rather artificially in the interest of model control, however, we have shown that the measurement term arises naturally through considering the effects of the information those measurements contain.  For this reason, we prefer to regard the measurement term as a guiding potential.  In the absence of the potential, we live in a state space restricted only by our model's degrees of freedom.  The introduction of the measurements guides us to a solution within a \textit{sub}space in which those particular measurements are possible.

\subsection{\textbf{Approximating the cost function}}

We now seek to simplify the cost function formulation for the purposes of calculation.

\subsubsection{The measurement term}

Regarding the measurement term, we make four assumptions:
\begin{itemize}
  \item The measurements taken at different times are independent of each other.  This permits us to write the CMI simply as: $log[P(\bm{x}(n)|\bm{y}(n))]$.  Or:
  \begin{equation*}
    A_0(\bm{X},\bm{Y}) = -\log[P(\bm{X}|\bm{Y})].
  \end{equation*}
  \item There may be an additional relation between the measurements and the state variables to which those measurements correspond, which can be expressed with the use of some transfer function $h_l$: $h_l(\bm{x}(n)) = y_l(n)$.
  \item For each of the $L$ measured state variables, we allow for a noise term $\theta_l$ at each timepoint, for each measurement $y_l$ that corresponds to a state variable $x_l$: $y_l(n) = h_l(\bm{x}(n)) + \theta_l(n)$.  In this case, then, $P(\bm{x}(n)|\bm{y}(n))$ is simply some function of $h(\bm{x}(n)) - \bm{y}(n)$ at each timepoint.
  \item The measurement noise has a Gaussian distribution.
\end{itemize}  
\noindent
Taking these assumptions, we arrive at:
\begin{align*}
  &\text{CMI}(\bm{x}(n),\bm{y}(n)|\bm{Y}(n-1)) \\
	&= -\sum_{l,k=1}^L (h_l(\bm{x}(n)) - y_l(n)) \frac{[R_m(n)]_{lk}}{2}(h_k(\bm{x}(n)) - y_k(n)),
\end{align*}
\noindent
where $R_m$ is the inverse covariance matrix of the measurements $y_l$.

\subsubsection{The model term}

We simplify the model term by assuming that the model may have errors, which will broaden the delta function in the expression noted earlier for the deterministic case.  If we assume that the distribution of errors is Gaussian, then $\delta^D(\bm{z})$ becomes: $\sqrt{\frac{\det R_f}{(2\pi)^D}}e^{[-\bm{z} \frac{R_f}{2} \bm{z}]}$, where $R_f$ is the inverse covariance matrix for the model's state variables.

\subsubsection{Cost function used for estimates in this paper}

Taking both approximations together, assuming that the transfer function $h_l$ is simply unity, and assuming that the minimizing path is independent of considerations of initial conditions, we obtain the cost function $A_0$ used in this paper, using a Hermite-Simpson integration method:
\end{multicols}
%\begin{widetext} 
\begin{gather*} \label{eq:costfunctioncomplete}
%\begin{split}
%A_0 = %\mathlarger{\sum}_n^{N-1} \mathlarger{\sum}_a^D \frac{R_a^f}{2}\left(x_a(n+1) - f_a(\bm{x}(n),\bm{p})\right)^2 \\
A_0 = \frac{R_f}{(N-1)D}\times\\ \sum_{n \in \{\text{odd}\}}^{N-2} \sum_{a=1}^D \left[ \left\{x_a(n+2) - x_a(n) - \frac{\delta t}{6} [F_a(\bm{x}(n),\bm{p}) + 4F_a(\bm{x}(n+1),\bm{p}) + F_a(\bm{x}(n+2),\bm{p})]\right\}^2  \right. \\
  + \left.\left\{ x_a(n+1) - \frac12 \left(x_a(n)+x_a(n+2)\right) - \frac{\delta t}{8} [F_a(\bm{x}(n),\bm{p}) - F_a(\bm{x}(n+2),\bm{p})]\right\}^2 \right] \\
  + \frac{R_m}{N_\text{meas}} \sum_j \sum_{l=1}^L (y_l(j) - x_l(j))^2.
%\end{split},
\end{gather*}
%\end{widetext}
\begin{multicols}{2}
\noindent
and we seek the path $\bm{X}^0 = \{\bm{x}(0),\ldots,\bm{x}(N),\bm{p}\}$ in state space on which $A_0$ attains a minimum value.  The two squared terms in the first double sum incorporate the model evolution of all $D$ state variables $x_a$. Of these, the first term in curly braces represents error in the first derivative (with respect to $t$) of the state variables, whereas the second term corresponds to error in the second derivative.  The outer summation in $n$ is taken over all odd-numbered grid points---discretized steps in $r$ that parameterize the model equations of motion.  The step-size $\delta t$ is defined as the distance between alternate grid points: $t(n+2) - t(n)$.  The inner summation in $a$ is taken over all $D$ state variables.

In the second term, the $N_{meas}$ coefficient is the number of timepoints at which measurements are made.

\section{Appendix 2: Details of the data assimilation procedure}
\label{appendix2}

\subsection{\textbf{Interface with Ipopt}}

Ipopt requires a user interface to discretize state space and calculate the model equations of motion, Jacobean, and Hessian matrices that are used in the minimization procedure.  A suite of Python codes was used to generate this interface; it is available here: https://github.com/yejingxin/minAone.

\subsection{\textbf{Choosing $R_f$/$R_m$ for best results}}

There exists no universal rule for choosing an optimal ratio of model and measurement weights.  An optimal value is model-dependent and must be identified via trial-and-error.  Generally, for many biophysical models of neurons, small neuronal networks, atmospheres, and chaotic Lorenz-63 and Lorenz-96 models, a value of $\beta$ between 10-20 is found to be ideal (private communications 2017).  The reader may compare this range to our identification of $\beta\in[13,15]$, which we found yielded the best results.

Poor results at the extremes ($R_m \gg R_f$ and $R_f \gg R_m$) are expected for any model, for the following reasons.  For low $R_f$, the model constraints are not yet sufficiently strict to require a converging solution.  For high $R_f$, the failure of solutions has at least two potential causes.  First, one encounters numerical problems with considering \lq\lq infinite\rq\rq\ model weight.  The problem is ill-conditioned when it involves a matrix whose elements are so large that the matrix is not invertible.  The optimizing solution may thus become overly sensitive to changes in the state vector.  Rounding error may render these solutions invalid.  A second possible cause is discretization error at high $R_f$.  In taking a discretized derivative, one retains only the first term in a Taylor series.  As the multiplicative factor grows, the higher-order terms - which are ignored - will become important. 

\section{Appendix 3: Biological motivation for the model explored in this paper, and mathematical formalism for sequenced neuronal activity}
\label{appendix3}

\subsection{\textbf{Biophysical significance of the model used in this paper}}
\begin{figure}[H]
  \centering
  \includegraphics[width=86mm]{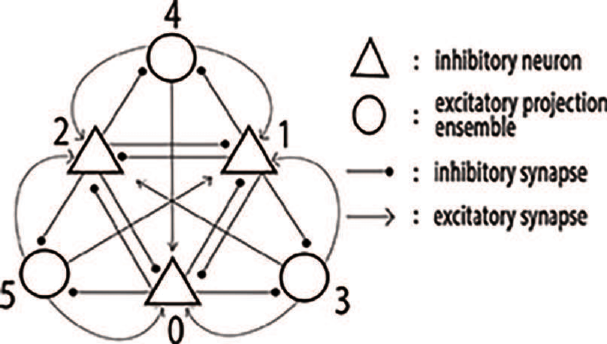}
%  \includegraphics[width=0.4\textheight]{Fig7.eps}
%\captionsetup{labelformat=empty}
\caption{{\bf A functional HVC unit, which, in Ref~\cite{armstrong2016model}, encoded a syllable of birdsong.}Three interneurons (triangles labeled 0, 1, and 2) are connected all-to-all.  Each interneuron connects directly to two out of the three excitatory $HVC_{RA}$ neurons (circles labeled 3, 4, and 5).}
\label{fig8}
\end{figure}

Figure~\ref{fig8} shows a structure of six neurons that Ref~\cite{armstrong2016model} considered to be a functional unit in HVC; here we shall refer to it as a functional HVC unit (FHU).  In an FHU, three interneurons (triangles: cells numbered 0, 1, and 2) are connected all-to-all, and each interneuron synapses directly to two of three HVC RA PNs (circles: cells numbered 3, 4, and 5).  (Feedback from the excitatory cells is also required for FHU functionality but is not an important consideration in this paper).  

\begin{figure}[H]
  \centering
  \includegraphics[width=0.6\textwidth]{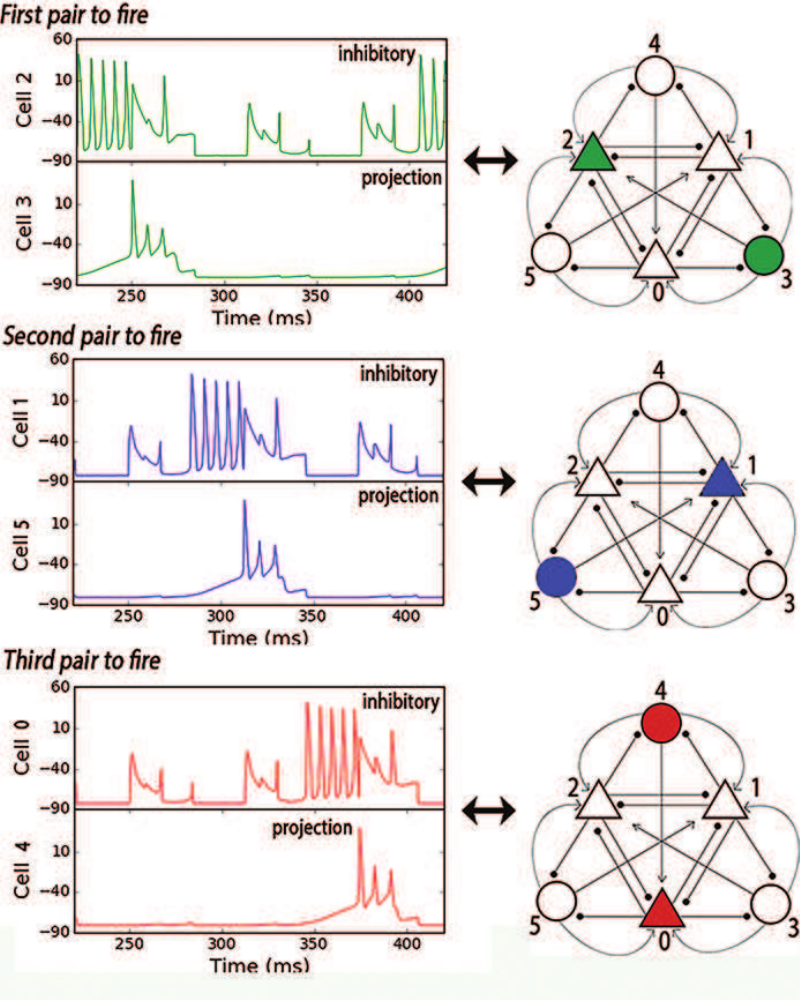}
%  \includegraphics[width=0.6\textwidth]{Fig8.eps}
%  \includegraphics[width=0.4\textheight]{Fig8.eps}
%\captionsetup{labelformat=empty}
\caption{{\bf A three-frame \lq\lq movie\rq\rq, representing sequential-bursting mode of a functional HVC (FHU).}  For a certain range of coupling strengths $g_{ij}$ among the interneurons, the interneurons may engage in a series of activations - each of which selects a particular $HVC_{RA}$ PN.  It is in this way that the observed sparse bursting~\cite{hahnloser2002ultra,kozhevnikov2007singing} may be mimicked.  (\textit{Reproduced from Ref~\cite{armstrong2016model}.})}
\label{fig9}
\end{figure}

The synaptic connections of interest are the all-to-all connections among the interneurons $g_{ij}$.  We shall focus on two modes of activity that may occur, given a low background of excitation.  For sufficiently low values of $g_{ij}$, all interneurons fire simultaneously, thereby suppressing all PNs (not shown); this mode captures the quiescence activity of the two neuronal populations.  For higher values of $g_{ij}$, the interneurons burst in a sequence, thereby effecting sequential firings of each PN (Figure~\ref{fig9}).  Ref~\cite{armstrong2016model} attributed a toggling between modes to a neuromodulatory process capable of rapidly increasing the inhibitory coupling strengths $g_{ij}$.  It is in this way that the observed sparse bursting of $HVC_{RA}$ PNs during song~\cite{hahnloser2002ultra,kozhevnikov2007singing} can be effected.  

\begin{figure}H]
  \centering
  \includegraphics[width=0.6\textheight]{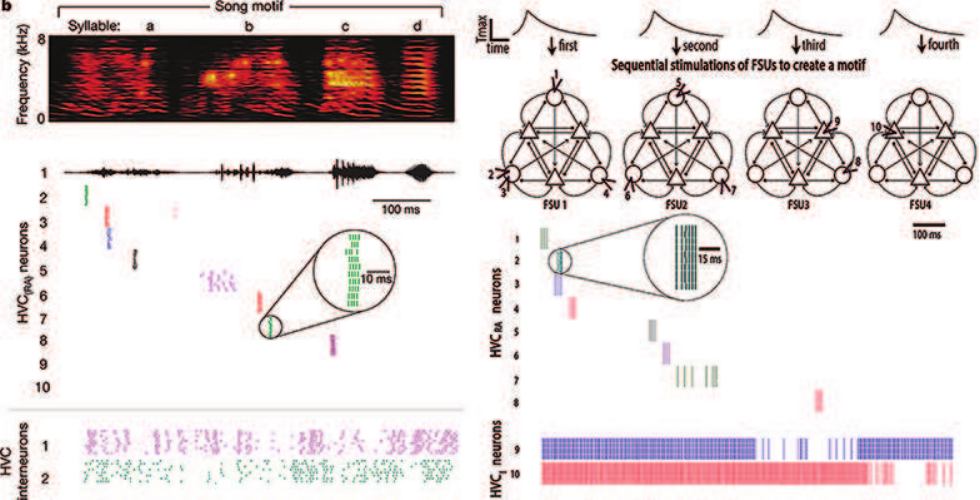}
%  \includegraphics[width=0.6\textheight]{Fig9.eps}
%\captionsetup{labelformat=empty}
\caption{{\bf Left: Raster plot of $HVC_{RA}$ PNs and $HVC$ interneurons observed during song~\cite{hahnloser2002ultra}.  Right: Simulated raster plot, using HVC model of Ref~\cite{armstrong2016model}.}  Note sparse bursting of $HVC_{RA}$ PNs and dense tonic spiking - with intermittent pauses - of HVC interneurons \textit{Reproduced from Ref~\cite{armstrong2016model}.}}
\label{fig10}
\end{figure}

This model reproduces basic qualitative features of HVC interneurons and RA-projecting PNs during song and during quiescence.  Figure~\ref{fig10}, reproduced from Ref~\cite{armstrong2016model}, compares the raster plot resulting from integrating the model equations of motion (right) with the experimental finding (left) of sparse $HVC_{RA}$ bursting during song~\cite{hahnloser2002ultra}.  The FHU structure also roughly captures the observed high rates of reciprocal connectivity between HVC interneuron and PN populations, and an observation that inhibition masks the activity of an excitatory population~\cite{kosche2015interplay}.
\end{multicols}
\begin{multicols}{2}
\subsection{\textbf{Formalism for competition among interneurons}}

The importance of inhibition for pattern generation was first formalized with the use of experimentally-obtained values of cell membrane conductances and inhibitory synaptic conductances to simulate cellular activity~\cite{hellgren1992computer,ekeberg1991computer, grillner1990cellular}.  Around that time, a collaboration between the experimental group of Selverston and a group led by Abarbanel and Rabinovich at the Institute for Nonlinear Dynamics - both at UC San Diego - sought to formalize basic operational principles of CPG activity.  

The first result of the UC collaboration was a dynamical model of the 14-neuron pyloric CPG~\cite{selverston2000reliable}.  Here they examined means by which a CPG may express predictable and reliable behavior even though comprised of neurons that, when isolated, may express chaotic dynamics.  They found that hyperpolarizing pulses regulated the neurons, while depolarizing pulses failed to do so.  From these studies there emerged a formalization of mutual inhibition among neurons: winnerless competition (WLC)~\cite{rabinovich2001dynamical}.

Within the WLC framework, neurons are represented by a collection of nodes that interact via competitive Lotka-Volterra-like dynamics.  For the case in which mutual inhibition results in sequential patterned activity among neurons, the nodes are saddle fixed points, and an orbit sequentially traverses limit cycles in the vicinity of each node.  A mode of simultaneous firing also exists, in which each node is rendered a stable fixed point.  The former and latter scenarios correspond to specific ranges of coupling strengths.  The WLC formalism captures fundamental features of CPG activity; hence it was harnessed as the basis for the model presented in Ref~\cite{armstrong2016model}.

\section{Appendix 4: Parameter values and estimates}
\label{appendix4}

\subsection{\textbf{Parameter estimates from pruning experiments}}

Table 5 lists parameter estimates for the case described in \textit{Results}, where the measurements are the voltage traces of all three cells, and where the maximum conductance of the $CaT$-type ion channel of Cell 0 is set to zero.  In Table 5 this value is underlined.

Table 6 lists parameter estimates for the case described in \textit{Results}, where the measurements are the voltage traces of all three cells, and where the synapse maximum conductances of four synapses - $g_{01}$, $g_{02}$, $g_{10}$, and $g_{12}$ - are set to zero.  In Table 6 these values are underlined.

\subsection{\textbf{Parameters held fixed throughout the data assimilation procedure}}

Tables 7 and 8 specify parameter values that were taken to be known throughout the DA procedure.  One exception - regarding the time constants $t_0$ and $t_1$ of each ion channel current - is described in \textit{Results}.  
\end{multicols}
\begin{table}[H]
%\begin{adjustwidth}{-2.25in}{0in} % Comment out/remove adjustwidth environment if table fits in text column.
\small
\centering
%\captionsetup{labelformat=empty}
\caption{Estimates of all parameters, for $g_{CaT,0}$ set to zero}
\setlength{\tabcolsep}{6pt}
\begin{tabular}{ l c c|c c|c c} \toprule
 \textit{Parameter} & \multicolumn{2}{c|}{Model with high $g_{ij}$ values} & \multicolumn{2}{c|}{Model with low $g_{ij}$ values} & Lower & Upper \\
  & \textit{estimate} & \textit{true} & \textit{estimate} & \textit{true} & bound & bound \\\midrule 
  \textit{$E_{01}$} & -82.95 & -83.0 & -83.0 & & -90.0 & 10.0 \\
  \textit{$E_{02}$} & -83.32 & -83.9 & -83.31& & \\
  \textit{$E_{10}$} & -82.67 & -82.7 & -82.59 & & \\  
  \textit{$E_{12}$} & -82.52 & -82.5 & -82.50 & & \\    
  \textit{$E_{20}$} & -83.08  & -83.2 & -83.35 & & \\
  \textit{$E_{21}$} & -82.87  & -82.9 & -82.88 & & \\\hline   
  \textit{$g_{L,0}$} & 2.90e-3  & 3.0e-3 & 2.98e-3 & & 9e-4 & 9e-2 \\    
  \textit{$g_{L,1}$} & 3.32e-3  & 3.3e-3 & 3.29e-3 & & & \\    
  \textit{$g_{L,2}$} & 2.84e-3  & 2.9e-3 & 2.88e-3 & & & \\\hline    
  \textit{$g_{Na,0}$} & 1.12 & 1.2 & 1.19 & & 0.2 & 1.8 \\    
  \textit{$g_{Na,1}$} & 0.97 & 1.0 & 0.97 & & & \\    
  \textit{$g_{Na,2}$} & 1.19 & 1.4 & 1.23 & & & \\\hline    
  \textit{$g_{K,0}$} & 0.190 & 0.2 & 0.196 & & 0.02 & 0.8 \\
  \textit{$g_{K,1}$} & 0.213 & 0.22 & 0.213 & & & \\    
  \textit{$g_{K,2}$} & 0.144 & 0.17 & 0.151 & & & \\\hline
  \textit{$g_{CaT,0}$} & 5.86e-7 & \underline{0.0} & 8.26e-8 & & 0.0 & e-2\\    
  \textit{$g_{CaT,1}$} & 1.17e-4 & 1.1e-4 & 1.09e-4 & & \\
  \textit{$g_{CaT,2}$} & 1.01e-4 & 0.9e-5 & 9.14e-5 & & \\\hline 
 \textit{$g_{01}$} & 0.249 & 0.25 & 0.0251 & 0.025 & 0.01 & 10.0 \\
 \textit{$g_{02}$} & 0.394 & 0.4 & 0.0396 & 0.040 & & \\
 \textit{$g_{10}$} & 0.281 & 0.28 & 0.0282 & 0.028 & & \\
 \textit{$g_{12}$} & 0.183 & 0.18 & 0.0179 & 0.018 & & \\
 \textit{$g_{20}$} & 0.216 & 0.21 & 0.0211 & 0.021 & &\\ 
 \textit{$g_{21}$} & 0.318 & 0.32 & 0.0321 & 0.032 & & \\\bottomrule
\end{tabular}
\newline \small{Estimates of all parameters, with $gCaT_0$ set to zero - once for the $g_{ij}$ underlying sequential bursting (the \lq\lq high\rq\rq\ values), and once for those underlying simultaneous firing (\lq\lq low\rq\rq).  The values that have been changed from the original design are underlined.  Unless noted, the true parameter values for the models with low versus high $g_{ij}$ values are identical.  Results are consistent over ten trials, and correspond to a $\beta$ value of 27. }\\\hrule
\label{table5}
%\end{adjustwidth}
\end{table}
\begin{table}[H]
%\begin{adjustwidth}{-2.25in}{0in} % Comment out/remove adjustwidth environment if table fits in text column.
\small
\centering
%\captionsetup{labelformat=empty}
\caption{Estimates of all parameters, with maximum conductances of four synapses set to zero}
\setlength{\tabcolsep}{6pt}
\begin{tabular}{ l c c|c c|c c} \toprule
 \textit{Parameter} & \multicolumn{2}{c|}{Model with high $g_{ij}$ values} & \multicolumn{2}{c|}{Model with low $g_{ij}$ values} & Lower & Upper \\
  & \textit{estimate} & \textit{true} & \textit{estimate} & \textit{true} & bound & bound \\\midrule 
  \textit{$E_{01}$} & N/A & & & & -90.0 & 10.0 \\
  \textit{$E_{02}$} & N/A & & & & \\
  \textit{$E_{10}$} & N/A & & & & \\  
  \textit{$E_{12}$} & N/A & & & & \\    
  \textit{$E_{20}$} & -83.17  & -83.2 & -83.15 & & \\
  \textit{$E_{21}$} & -82.87  & -82.9 & -82.86 & & \\\hline   
  \textit{$g_{L,0}$} & 2.99e-3  & 3.0e-3 & 2.99e-3 & & 9e-4 & 9e-2 \\    
  \textit{$g_{L,1}$} & 3.30e-3  & 3.3e-3 & 3.29e-3 & & & \\    
  \textit{$g_{L,2}$} & 2.84e-3  & 2.9e-3 & 2.88e-3 & & & \\\hline    
  \textit{$g_{Na,0}$} & 1.16 & 1.2 & 1.16 & & 0.2 & 1.8 \\    
  \textit{$g_{Na,1}$} & 1.02 & 1.0 & 0.98 & & & \\    
  \textit{$g_{Na,2}$} & 1.26 & 1.4 & 1.20 & & & \\\hline    
  \textit{$g_{K,0}$} & 0.194 & 0.2 & 0.193 & & 0.02 & 0.8 \\
  \textit{$g_{K,1}$} & 0.223 & 0.22 & 0.215 & & & \\    
  \textit{$g_{K,2}$} & 0.150 & 0.17 & 0.147 & & & \\\hline
  \textit{$g_{CaT,0}$} & 9.92e-5 & 1.0e-4 & 9.76e-5 & & e-5 & e-2\\    
  \textit{$g_{CaT,1}$} & 1.05e-4 & 1.1e-4 & 1.06e-4 & & \\
  \textit{$g_{CaT,2}$} & 1.11e-4 & 0.9e-5 & 9.14e-5 & & \\\hline 
 \textit{$g_{01}$} & 1.0e-4 & \underline{0.0} & 1.4e-3 & & 0.0 & 10.0 \\
 \textit{$g_{02}$} & 0.0 & \underline{0.0} & 3.7e-5 & & & \\
 \textit{$g_{10}$} & 8.84e-6 & \underline{0.0} & 4.8e-6 & & & \\
 \textit{$g_{12}$} & 0.0 & \underline{0.0} & 0.0 & & & \\
 \textit{$g_{20}$} & 0.193 & 0.21 & 0.021 & 0.021 & &\\ 
 \textit{$g_{21}$} & 0.350 & 0.32 & 0.032 & 0.032 & & \\\bottomrule
\end{tabular}
\newline \small{Estimates of all parameters, with four of the $g_{ij}$ set to zero - once using the $g_{ij}$ underlying sequential bursting for the other two (non-zero) $g_{ij}$ (the \lq\lq high\rq\rq\ values), and once using those underlying simultaneous firing (\lq\lq low\rq\rq).  The values that have been changed from the original design are underlined.  Unless noted, the true parameter values for the models with low versus high $g_{ij}$ values are identical.  Results are consistent over ten trials, and correspond to a $\beta$ value of 27. }\\\hrule
\label{table6}
\vspace{10cm}
%\end{adjustwidth}
\end{table}
\begin{multicols}{2}
\begin{table}[H]
%\begin{adjustwidth}{-2.25in}{0in} % Comment out/remove adjustwidth environment if table fits in text column.
\small
\centering
%\captionsetup{labelformat=empty}
\caption{Cellular parameters taken to be known}
\setlength{\tabcolsep}{4pt}
\begin{tabular}{ l c c c c} \toprule
 \textit{Parameter} & Cell 0 & Cell 1 & Cell 2 & Unit\\\midrule 
 \textit{$E_{L}$} & -70.0 & -65.5 & -70.5 & mV\\
 \textit{$E_{Na}$} & 50.0 & 50.5 & 49.5 & mV\\
 \textit{$E_{K}$} & -77.0 & -76.5 & -76.8 & mV\\
 \textit{$\theta_{m}$} & -40.0 & -40.5 & -40.2 & mV\\
 \textit{$\sigma_{m}$} & 16.0 & 15.5 & 16.5 & mV\\
 \textit{$t_{0,m}$} & 0.1 & 0.11 & 0.09 & ms\\
 \textit{$t_{1,m}$} & 0.4 & 0.41 & 0.43 & ms\\
 \textit{$\theta_{h}$} & -60.0 & -59.5 & -59.8 & mV\\
 \textit{$\sigma_{h}$} & -16.0 & -15.6 & -16.6 & mV\\
 \textit{$t_{0,h}$} & 1.0 & 1.01 & 1.02 & ms\\
 \textit{$t_{1,h}$} & 7.0 & 6.9 & 7.2 & ms\\
 \textit{$\theta_{n}$} & -60.0 & -54.5 & -55.5 & mV \\
 \textit{$\sigma_{n}$} & 25.0 & 24.5 & 24.3 & mV\\
 \textit{$t_{0,n}$} & 1.0 & 0.99 & 0.97 & ms\\
 \textit{$t_{1,n}$} & 5.0 & 4.9 & 5.1 & ms\\
 \textit{$\theta_{a}$} & -70.0 & -70.5 & -69.0 & mV\\
 \textit{$\sigma_{a}$} & 10.0 & 11.0 & 9.0 & mV\\
 \textit{$t_{0,a}$} & 1.0 & 1.1 & 0.9 & ms\\
 \textit{$t_{1,a}$} & 5.0 & 5.21 & 5.19 & ms\\
 \textit{$\theta_{b}$} & -65.0 & -64.5 & -65.2 & mV\\
 \textit{$\sigma_{b}$} & -10.0 & -11.0 & -9.2 & mV\\
 \textit{$t_{0,b}$} & 1.0 & 1.1 & 0.9 & ms\\
 \textit{$t_{1,b}$} & 100.0 & 100.1 & 99.0 & ms\\
 \textit{$\phi$} & 0.06 & 0.05 & 0.07 & $\mu$M/ms/nA\\
 \textit{$Ca_0$} & 0.2 & 0.21 & 0.19 & $\mu$M\\
 \textit{$C$} & 0.01 & 0.011 & 0.009 & $\mu$F\\
 \textit{$\tau_{Ca}$} & 10.0 & 13.0 & 9.0 & ms\\
 \textit{$T$} & 290 & & & K \\
 \textit{$Ca_{ext}$} & 2500 & & & $\mu$M  \\\bottomrule
\end{tabular}
\label{table7}
%\end{adjustwidth}
\end{table}
\begin{table}[H]
%\begin{adjustwidth}{-2.25in}{0in} % Comment out/remove adjustwidth environment if table fits in text column.
\small
\centering
%\captionsetup{labelformat=empty}
\caption{Synapse parameters taken to be known}
\setlength{\tabcolsep}{12pt}
\begin{tabular}{ l c c c c} \toprule
 \textit{Parameter} & Cell 0 & Cell 1 & Cell 2 & Unit\\\midrule 
 \textit{$T_{max}$} & 1.5 & 1.49 & 1.51 & mM\\
 \textit{$V_{p}$} & 2.0 & 2.01 & 2.03 & mV \\
 \textit{$K_{p}$} & 5.0 & 5.01 & 4.8 & mV\\\bottomrule
\end{tabular}
\label{table8}
%\end{adjustwidth}
\end{table}

\end{multicols}
%\nolinenumbers

% Either type in your references using
% \begin{thebibliography}{}
% \bibitem{}
% Text
% \end{thebibliography}
%
% or
%
% Compile your BiBTeX database using our plos2015.bst
% style file and paste the contents of your .bbl file
% here. See http://journals.plos.org/plosone/s/latex for 
% step-by-step instructions.
% 
%\bibliographystyle{acm}
\bibliographystyle{unsrt}
\bibliography{refs_Rewrite18aug}

\end{document}